\documentclass[11pt,english]{article}
\pdfoutput=1
\usepackage{jheppub}
\usepackage{lscape}
\usepackage{amsmath}
\usepackage{slashed}
\usepackage{mathtools}
\usepackage{cleveref}
\usepackage{tcolorbox}

\usepackage{subcaption}
\usepackage{amssymb,tikz}
\usetikzlibrary{positioning}
  \def\gn#1#2{{$\href{http://groupnames.org/\#?#1}{#2}$}}
\def\gn#1#2{$#2$}  
\tikzset{sgplattice/.style={inner sep=1pt,norm/.style={red!50!blue},char/.style={blue!50!black},
  lin/.style={black!50}},cnj/.style={black!50,yshift=-2.5pt,left=-1pt of #1,scale=0.5,fill=white}}
\usetikzlibrary{shapes.geometric, arrows}

\usetikzlibrary{decorations.markings}
\tikzstyle{startstop} = [rectangle, rounded corners, 
minimum width=3cm, 
minimum height=1cm,
text centered, 
draw=black, 
fill=red!30]

\tikzstyle{io} = [trapezium, 
trapezium stretches=true, 
trapezium left angle=70, 
trapezium right angle=110, 
minimum width=3cm, 
minimum height=1cm, text centered, 
draw=black, fill=blue!30]

\tikzstyle{process} = [rectangle, 
minimum width=3cm, 
minimum height=1cm, 
text centered, 
text width=3cm, 
draw=black, 
fill=orange!30]

\tikzstyle{decision} = [diamond, 
minimum width=3cm, 
minimum height=1cm, 
text centered, 
draw=black, 
fill=green!30]
\tikzstyle{arrow} = [thick,->,>=stealth]

\def\-{\hphantom{-}}
\def\ov{\overline}
\def\s2{\frac{1}{\sqrt2}}

\def\be{\begin{equation}}
\def\ee{\end{equation}}
\def\bea{\begin{align}}
\def\eea{\end{align}}
\def\beqa{\begin{eqnarray}}
\def\eeqa{\end{eqnarray}}

\def\mg{m_{3/2}}
\def\mg2{m^2_{3/2}}

\def\Dsl{\,\raise.15ex\hbox{/}\mkern-13.5mu D} 

\newcommand{\eq}[1]{\begin{equation}
                     \begin{split} #1 \end{split}
                     \end{equation}}

\begin{document}

\title{Galois groups of uplifted de Sitter vacua}

\author[a,b]{Cesar Damian} 
\author[c]{and Oscar Loaiza-Brito}

\affiliation[a]{Departamento de Ingenier\'ia Mec\'anica, Universidad de Guanajuato, Carretera Salamanca-Valle de Santiago Km 3.5+1.8 Comunidad de Palo Blanco, Salamanca, Mexico.}
\affiliation[b]{Department of Mechanical Engineering, Virginia Tech, Blacksburg, VA 24061, USA.}
\affiliation[c]{Departamento de F\'isica, Universidad de Guanajuato, Loma del Bosque No. 103 Col. Lomas del Campestre C.P 37150 Leon, Guanajuato, Mexico.}


\date{\today}

\abstract{We compute the Galois group of a polynomial whose roots are determined by the critical points of a scalar potential in type IIB compactifications.
We focus our study on certain perturbative models where it is feasible to construct a de Sitter vacuum within the effective theory by introducing non-geometric fluxes, D-branes or non-BPS states. Our findings
clearly show that all de Sitter vacua derived from lifting AdS stable vacua are associated with an unsolvable Galois group. This suggests a deeper connection between the fundamental principles of Galois theory and its applications in the construction of dS vacua.
}
\arxivnumber{}

\keywords{Galois theory, dS vacua, Swampland}

\maketitle

%
%

\section{Introduction}
The Swampland program is a research program in string theory that aims to identify which effective quantum field theories can be consistently embedded into a theory of quantum gravity, such as string theory, and which cannot. It is motivated by the observation that many seemingly consistent effective quantum field theories appear to be incompatible with some expected and verified principles of quantum gravity, which is studied by general scenarios in string theory (generally involving black hole physics). The program aims to identify and understand these inconsistencies and use them to constrain the space of quantum field theories with an ultraviolet completion in string theory \cite{Palti:2019pca,vanBeest:2021lhn,Grana:2021zvf,Agmon:2022thq}.\\

In the last few years, important insights into string theory and quantum gravity have been identified by proposing several different conjectures that appear to be correlated nonetheless. This has stimulated research into the connections between string theory, cosmology, and other areas of physics. One of these is the so-called Refined de Sitter Conjecture (RdSC)\cite{Garg:2018reu,Ooguri:2018wrx}, which states that the existence of stable de Sitter vacua is excluded at least in asymptotic regions of the moduli space. However, unlike well-established conjectures such as the weak gravity or the absence of global symmetries, the RdSC is at the moment a proposal based mainly on the enormous difficulty to construct stable de Sitter vacua from string compactifications rather than the consequence of a more fundamental constraint in a quantum gravity theory, although there are some strong grounds to consider its validity \cite{Andriot:2018mav,Blumenhagen:2019kqm,Seo:2019mfk,Brandenberger:2020oav,Damian2023some}.\\

One of the most useful techniques to construct dS vacuum is to uplift some direction which is previously stabilized at some non-positive value, although it is also possible to uplift a runaway direction. The most well-known scenarios are the KKLT and the LVS, where the stability of the de Sitter vacuum has been extensively studied in the last twenty years (for a recent overview see \cite{Leontaris:2023obe}). More simple scenarios, such as toroidal compactifications in the presence of fluxes, also exhibit the presence of apparently stable de Sitter minima in some directions of the scalar potential which are compromised by topological constraints or by the inclusion of exotic fluxes \cite{Damian:2013dq,Damian:2013dwa,Blumenhagen:2015kja,Damian:2018tlf, Plauschinn:2018wbo, Andriot:2019wrs,  Shukla:2019wfo, Shukla:2022srx}. However, these scenarios offer the possibility to study a relationship between the presence of de Sitter minima and algebraic properties of the scalar potential and its roots since in these cases they are polynomials of order cubic at most on the moduli. In this context, we are interested in searching for some inherent algebraic characteristics of a polynomial scalar potential related to scenarios where one or more directions with some negative or null minima are uplifted to a positive one.\\

For that, we examine four different scenarios in which it is possible to uplift some directions to stable de Sitter minima by adding extra structure such as non-geometric fluxes, non-BPS states, or by relaying in a perturbative polynomial truncation of the superpotential related to a Calabi-Yau manifold with large complex structure as reported in \cite{Coudarchet:2022fcl}. In all of these cases, the existence of such minima has been numerically demonstrated.\\

In this work, we show that there is a common algebraic characteristic associated with these scenarios, specifically that the Galois group of the polynomial scalar potential is unsolvable when uplifted to a de Sitter minima, but solvable otherwise. We believe that this indicates a deeper relationship between the algebraic properties of the scalar potential and the existence of apparently stable de Sitter vacua. Since all of these vacua are related to scenarios for which their ultraviolet completion is not clear, one could say that a de Sitter vacuum constructed perturbatively possesses a scalar potential with an unsolvable Galois group. \\

Our work is organized as follows: In section 2 we provide some basics on the Galois theory by covering the essentials to construct the Galois group. Sections 3 to 6 are focused on the study of different cases in which we can uplift some directions to a positively valued minimum. In all of them we find that previous to the uplifting, the associated Galois group is solvable but becomes unsolvable once we manage to uplift the minimum to a dS one. We show our conclusions and final comments in Section 7, while in the Appendix we present some techniques that could be useful to the reader.

\section{A brief overview on Galois group}
Our goal in this work is to compute the Galois group for different scenarios involving a polynomial scalar potential and to look for a relation between the different stable vacua we can construct and the corresponding Galois group. To that end, we shall start with a very brief overview of Galois Theory and its importance in the existence of different vacua.\\

\subsection{The Galois group}
The Galois group of a polynomial is a fundamental concept in algebraic number theory as well as algebraic geometry. Algebraically the Galois group is the group of automorphisms of the field generated by the extension of the rational field by including the roots of the polynomial, such that the field operations are preserved. This group provides important information on the structure and relation of the roots of the polynomial.\\

The field extension is a larger field that usually contains as the base field the rationals. For instance, the biggest field extension over the reals is the field of complex numbers.  In the same manner, the field of algebraic numbers is an extension of rational numbers and encodes information regarding the symmetries of the field extension in its associated Galois group. The computation of the Galios group of a field extension is carried out by finding all the automorphisms of the extension that left fixed the base field and indeed is a bijective map that preserves the field operations. \\

 For example, for $n\in\mathbb{Z}^+$, the field extension generated by the polynomial $p(x)=x^2-n$ over the field of rational numbers contains the square root of $n$. The automorphisms, $\sigma$ that fix the rational numbers must take $\sqrt{n}$  to another root of the polynomial. In this case, $-\sqrt{n}$. Thus, the automorphism that takes $\sigma ( \sqrt{n} )$ to $-\sqrt{n}$ is one element of the Galois group.\\
 
Since the computation of the Galois group could be a non-familiar calculation in physics, we give a short and very summarized set of steps  we have used to compute it, corresponding to  an implementation of the Stauduhar method to determine the Galois group \cite{awtrey2017determining, wildsmithcontribution}:
 \begin{enumerate}
\item Once we have constructed the associated polynomial to some moduli field from the scalar potential, we check whether the polynomial is irreducible over its base field. If it is reducible, then we can factor it into irreducible factors and compute the Galois group of each irreducible factor separately.
\item Next, we need to find all the roots of every irreducible polynomial, which are the elements of the splitting field. We can use algebraic or numerical methods to find the roots, and the roots are ordered in an arbitrary manner.
\item Let us say that $\text{Gal} ~p$ is the Galois group of the polynomial $p(x)$ with respect to the initial arbitrary ordering of the corresponding roots.  One way to figure out what $\text{Gal} ~p$ is, consists on identify possible subgroups $M$. To determine if  $M$ is a subgroup of $\Gamma$ one can follow a set of steps as described in Ref. \cite{stauduhar1973determination,GEISSLER2000653}. Broadly speaking one requires to compute the resolvent of the polynomial.
The resolvent is a monic polynomial $p$ with integer coefficients to which one can relate some trimmed invariants, as the discriminant (to be defined later on). These invariants are tested for integer values: if no integer values are found then it is concluded that $\Gamma $ are not contained in any conjugates of $M$, and similar resolvents are computed for other conjugacy classes of the maximal transitive subgroup of S$_n$. On the other hand, if an integer root is found for the coset representative $\pi_i$, this is used to re-order the roots, and then $\text{Gal}~p$ is a subgroup $\pi p  \pi^{-1}$. 
\item This procedure is repeated over a so-called {\it decision tree} to check all the possibilities to uniquely determine the Galois group \cite{wildsmithcontribution}. In the cases we shall explore, we use some results directly from those reported in references \cite{wildsmithcontribution,SUTHERLAND201573}.
\end{enumerate}

A Galois group is {\it unsolvable} if it is not a solvable group. A solvable group is a group that can be solved using a sequence of abelian groups and their extensions.
The fundamental theorem of Galois theory implies that the solvability of the Galois group is directly related to the fact that a polynomial can be solved for radials, namely, whether the roots of the polynomial can be expressed by elementary operations of a field, such as multiplication, division, and square roots. Thus, the Abel-Ruffini theorem, states that polynomials of degree five and higher are not solvable for radicals, always that its Galois group is not solvable \cite{stewart2022galois}. In all the cases we shall study, the polynomial related to dS vacua are of degree greater than 5.\\


Currently, it is possible to find several computational tools as well as algorithms to compute the Galois theory. Some common tools employed are specialized software such as Mathematica, Sagemath, and Magma which have built-in functions for computing Galois groups, which allows one to determine the Galois group by determining the action of the permutation group acting on the roots \footnote{ Since we are mainly interested in transitive groups and we shall relate the groups given in the decision tree reported in \cite{wildsmithcontribution}, with the common notation given for transitive groups of degree less than 32, as $n T m$ where $n$  is the degree of the group and $m$ is the number of the group in the classification \cite{transitivegroups} and take its corresponding group alias.}.\\

In the subsequent sections, we will explore four different scenarios where the uplifting mechanism from a stable AdS vacuum to a dS one is constructed by the addition of anti $D3$-branes, non-geometric fluxes, and non-BPS states. Through these examples, we shall conclude that all such uplifting scenarios are characterized by a lack of analyticity in the process and in consequence are linked to an unsolvable Galois group.\\

\section{Example 1:  A model without complex structure}
Let us consider a IIB string compactification on a CY with a single K\"ahler modulus and without complex structure in the presence of a NS-NS flux $h$, a RR flux $f$ and a non-geometric flux $q$ (model A in \cite{Blumenhagen:2015kja}). This emulates a six-dimensional torus with a frozen complex structure. The superpotential is given by
\eq{
W = i \tilde f + i h S + i q T
}
with K\"ahler moduli
\eq{
K = - 3 \log \left( T + \bar T \right) - \log \left( S+ \bar S \right)
}
Hence, the scalar potential takes the form
\eq{
V=\frac{c^2 h^2}{16 s \tau ^3}-\frac{c h \rho  q}{8 s \tau ^3}-\frac{\tilde f h}{8 \tau ^3}-\frac{\tilde f q}{8 s \tau ^2}+\frac{\tilde f^2}{16 s \tau ^3}-\frac{3 h q}{8 \tau ^2}+\frac{h^2 s}{16 \tau ^3}+\frac{\rho ^2 q^2}{16 s \tau ^3}-\frac{5 q^2}{48 s \tau }
}
with $S= s+ i c$ and $T = \tau+ i \rho$. In our next step, we shall compute the Galois group related to the AdS minimum.

\subsection{AdS free tachyon}
The critical points of V are located at the zeros of the two polynomial equations given by $\partial_{s,\tau} V= 0$ with $\partial_{c,\rho} V$ identically zero at $c =\rho =0$. The zeros are located at 

\eq{
\left( s, \tau \right) = \left[ \left( -\frac{\tilde f}{2h}, - \frac{3}{2}\frac{\tilde f}{q}\right) , \left(  -\frac{\tilde f}{h},  - \frac{6}{5}\frac{\tilde f}{q}  \right) , \left( \frac{\tilde f}{8h}, \frac{3}{8}\frac{\tilde f}{q} \right) \right]
}
where, as shown in \cite{Blumenhagen:2015kja}, only the second solution is free of tachyons. This vacuum  is AdS with an energy value given by
\eq{
V_{\text{min}} = - \frac{5^2}{2^3 \cdot 3^3 }\frac{q^3 h}{\tilde f^2} \,,
}
and the mass spectrum of the form
\eq{
m^2{_i} = \alpha_i \frac{h q^3}{\tilde f^2}
}
for $i=s,\tau$ and for $\alpha_i$ a numerical value greater than 1. By restricting to $\tau = -\frac{6}{5} \frac{\tilde f}{q}$, the zeros of the axiodilaton $s$ are elements of the splitting field of the polynomial 
\eq{
p(s) = (h s+\tilde f) (2 h s+\tilde f) (8 h s- \tilde f) = 0,
}
for $f, h, q \in \mathbb{Z}$ as expected from Dirac quantization. Thus, the polynomial already splits in the field $\mathbb{Q}$ and the set of automorphisms only contains the identity element.  Then the Galois group is trivial:
\eq{
\text{Gal}\, p( s) = e.
}

\begin{figure}[htbp]
   \centering
   \begin{tikzpicture}
   \draw[->,ultra thick] (-5,0)--(2.5,0) node[right]{Re $s$};
   \draw[->,ultra thick] (-2,-1)--(-2,5) node[above]{Im $s$};
   
   \draw[ultra thick, blue, dashed, decoration={markings, mark=at position 0.625 with {\arrow{>}}},postaction={decorate}] (-3,0.5) circle (0.5);
         \draw[ultra thick, blue, dashed, decoration={markings, mark=at position 0.625 with {\arrow{>}}},postaction={decorate}] (-1.0,0.5) circle (0.5);
          \draw[ultra thick,blue, decoration={markings, mark=at position 0.625 with {\arrow{>}}},postaction={decorate}] (1.5,0.5) circle (0.5);
        \node at (-3,1.2) {$e$};
        \node at (-3,-0.5) {$\frac{\tilde f}{8h}$};
         \node at (-1,1.2) {$e$};
         \node at (-1.0,-0.5) {$-\frac{\tilde f}{2h}$};
        \node at (1.5,1.2) {$e$};
        \node at (1.5,-0.5) {$-\frac{\tilde f}{h}$};
   \end{tikzpicture}
   
   \caption{Roots of p(s) and the trivial automorphisms. Dashed lines represent the automorphism on non-physical roots, while the continuous one is the automorphism to the unique physical root.}
   \label{fig:1}
\end{figure}
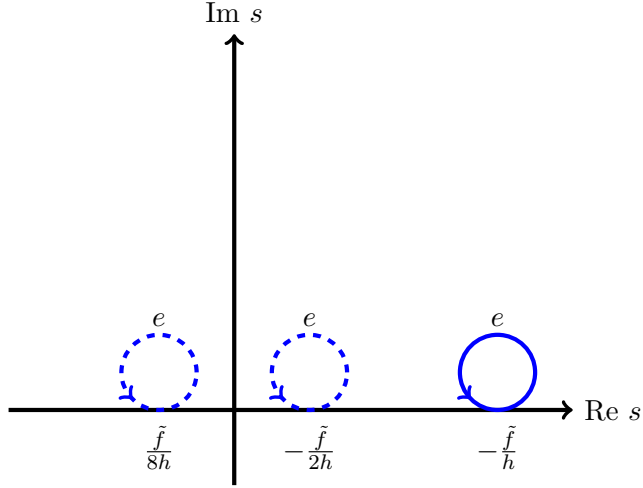

This implies that all vacua are isolated to each other by any automorphism and it is not possible to connect a vacuum to any other from an algebraic point of view.  In Figure \ref{fig:1} the dashed circles correspond to the automorphisms of the vacua with a tachyonic direction, whereas the solid one corresponds to the non-tachyon case. \\

Notice also that
the physical vacuum requires that
\eq{
 \tilde f < 0 \, \quad  \text{and} \quad h, q >0 \,,
}
with a $D3$-brane tadpole contribution given by
\eq{
N_{D3}^{\text{flux}}= \tilde f h.
}
Thus, since the automorphisms are trivial,  the tadpole remains unchanged, implying that $N_{D3}^{\text{flux}}< 0$ and in turn requires the presence of   D3 branes (or positively charged orientifold planes).\\

Notice that we have selected the construction of the Galois group for the polynomial generated by the roots of $s$ while evaluating at some fixed value of $\tau$. Although we can perform the same procedure for $\tau$ and in this case obtain that $\text{Gal}~ p(\tau)=e$, in the subsequent examples we shall restrict our study to the polynomial generated by the critical values on the $s$-direction.\\

\subsection{Uplift to dS}
Let us now consider the case where an $\overline{D3}$ is added to the scalar potential to uplift the free-tachyon vacua into a Minkowski or dS one. The anti $D3$-brane contribution to the scalar potential is given by
\eq{
V_{D3} = \frac{A_{\overline{D3}}}{\mathcal{V}^{4/3}} \,,
}
with ${\cal V} = (2\tau)^{3/2}$. To find the critical points, we first consider the set of algebraic equations
\eq{
\partial_s V = 0 \,, \partial_\tau V = 0 \,, V = \Lambda ,
} 
where the condition $V = \Lambda$ can be solved for  $A_{\overline{D3}}$ being a function of $\tau$ and $s$. Thus in the conditions of $\partial_s V = \partial_\tau V = 0$ the dependence of $A_{\overline{D3}}$ is removed resulting in two polynomials in $s$ and $\tau$. The condition $\partial_s V = 0$ can be solved for $\tau$ since is a polynomial of degree 2, and these roots are employed to leave $\partial_\tau V = 0$ as a polynomial depending only on $s$. The roots of $s$ now are contained in the splitting field defined by the irreducible polynomial over $\mathbb{Q}$,
\begin{align}
&f(s) =  25 \tilde f^2 q_1^6-576 \tilde f^3 \Lambda  q^3 s+\left( -864 \tilde f^4 \Lambda ^2 -180 \tilde f^2 h \Lambda  q^3 \right) s^2+  \\ \nonumber
&\left(-1728 \tilde f^3 h \Lambda ^2 +180 \tilde f h^2 \Lambda  q_1^3 \right) s^3 + 1728 \tilde f h^3 \Lambda ^2 s^5+864 h^4 \Lambda ^2 s^6  =0.
\end{align}
This non-monic polynomial with integer coefficients can be transformed into a monic polynomial $p(\tilde s)$ with discriminant (see Appendix for details and definitions)
\begin{align}
\text{disc}\, p(\tilde s) &= -a \tilde f^{10} q^{12} h^{94} \Lambda^{50} \sum_{i=0}^7 a_i \tilde f^{2 i} h^{6-i} \Lambda^{i} q^{18-3 i} 
\end{align}
with
\eq{
a&= 2^{122} \cdot 2^{78}\\
a_0 &= 3^3 \cdot 5^{13} \\
a_1 &= 2^4 \cdot 3^{2}\cdot 5^9\cdot 11\cdot 773\\
a_2 &= 2^7 \cdot 5^5 \cdot 114296701\\
a_3 &= 2^{10}\cdot 3^{3}\cdot 5^{4}\cdot  7976063 \\
a_4 &= 2^{13}\cdot 3^{9}\cdot 5^2\cdot 1770757 \\
a_5 &= 2^{20}\cdot 3^{9}\cdot 11^1 \cdot 35869 \\  
a_6 &= 2^{25}\cdot 3^{12}\cdot 11^2 \\  
}
Clearly, $\text{disc}\, p(\tilde s) $ is not a perfect square. This rules out the alternating group $\text{A}_6$, as well as its subgroups, to be the Galois group associated to $p(s)$. \\

Besides, for $\Lambda =0$ corresponding to a Minkowski vacuum, the discriminant becomes zero, indicating that there is a degeneracy on different roots of $p(\tilde s)$.  This degeneracy is broken once $\Lambda$ takes non-zero values. \\

Therefore, for $\Lambda\ne 0$,  to obtain the Galois group of $p (\tilde s)$ we check the chain
\eq{
\left[ (S_3)^2 \right]^2 2  \rightarrow  2^3 \left[ S_3 \right]  \rightarrow \sqrt{\Delta p(\tilde s)} \in \mathbb{Z} \rightarrow PGL(2,5)  \rightarrow {\bf S_6} \,,
}
which has a trimmed invariant polynomial over $S_6$ given by
\eq{
\left( \tilde s_1 \tilde s_2 +\tilde s_3 \tilde s_5 + \tilde s_4 \tilde s_6 \right) \left( \tilde s_1 \tilde s_3+ \tilde s_4 \tilde s_5 +\tilde s_2 \tilde s_6 \right) \cdot \\
 \left( \tilde s_3 \tilde s_4 +\tilde s_1 \tilde s_6 +\tilde s_2 \tilde s_5 \right) \left( \tilde s_1 \tilde s_5 + \tilde s_2 \tilde s_4 +\tilde s_3 \tilde s_6 \right) \left( \tilde s_1 \tilde s_4 +\tilde s_2 \tilde s_3+\tilde  s_5 \tilde s_6 \right) = 0\,,
}
which representative cosets listed in \cite{stauduhar1973determination}. Altogether leads us to the conclusion that the Galois group for $p(s)$ with $\Lambda \ne 0$ is
\eq{
\text{Gal}\, p(s) = \mathbf{ S_6},
}
which is non-solvable and transitive.  By applying the Complex Conjugate Root Theorem, we can deduce that the polynomial $p(s)$ includes three pairs of complex conjugate roots. Consequently, the physical vacuum state is contained in at least two real roots, which can be connected by automorphisms in the case that is absolute value is equal for the two roots. However, numerical inspection locates the two real positive roots in different absolute values, indicating that the real root remains isolated from any automorphism of the Gal $(p(s))$.\\

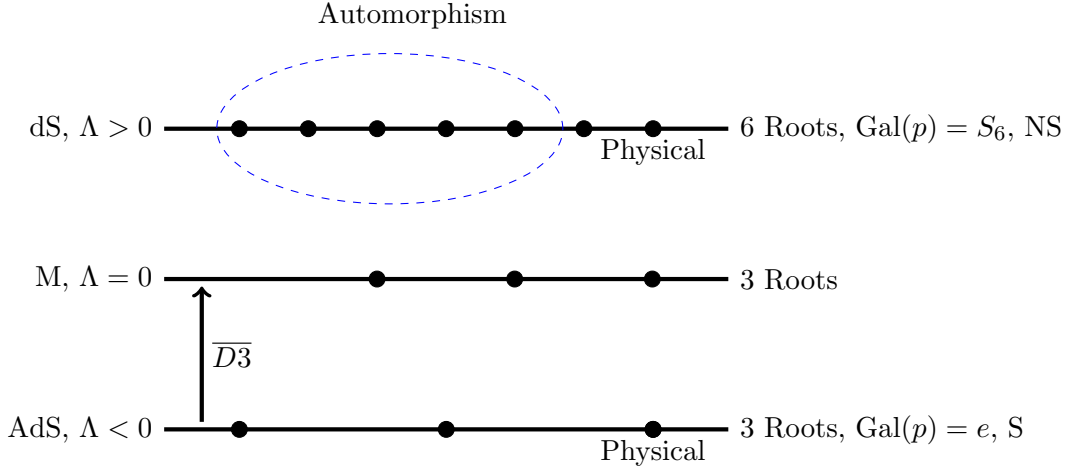
\begin{figure}[htbp]
   \centering
   \begin{tikzpicture}
	\draw[-,ultra thick] (-5,0) node[left]{AdS, $\Lambda < 0$} --(2.5,0) node[right]{3 Roots,  Gal$(p)=e$, S};
	\draw[-,ultra thick] (-5,2) node[left]{M, $\Lambda = 0$} --(2.5,2) node[right]{3 Roots};
	\draw[-,ultra thick] (-5,4) node[left]{dS, $\Lambda > 0$} --(2.5,4) node[right]{6 Roots,  Gal$(p)=S_6$, NS};
	\draw[->,ultra thick] (-4.5,0.1)  --(-4.5,1.9) node at (-4.1,1.0) {$\overline{D3}$};
	\filldraw (1.5,4) circle (3pt) node[below]{Physical};
	\draw[blue,dashed] (-2.0,4) ellipse (2.3cm and 1cm);
	\node at (-1.7,5.5) {Automorphism};   
	\foreach \x in {0,1,2} { \filldraw (-4+2.75*\x,0) circle (3pt);}  \filldraw (1.5,0) circle (3pt) node[below]{Physical};
	\foreach \x in {1,2,3} { \filldraw (-4+1.83*\x,2) circle (3pt);}
	\foreach \x in {0,1,2,3,4,5} { \filldraw (-4+0.916*\x,4) circle (3pt);}
   \end{tikzpicture}
   \caption{There are 3 different stable AdS vacua but only one is physical. The Galois group is the identity $e$. After uplifting to dS by the addition of anti $D3$-branes, the number of roots increases to 6. Four of them are connected through automorphisms of $S_6$ and there are two roots isolated. S=Solvable, NS=Non-solvable.}
   \label{fig:r1}
\end{figure}

We also notice by a numerical inspection, that the mass of the moduli requires that the sign of the RR flux $\tilde f$ is opposite to the RR flux in an AdS vacuum.  Namely, the free tachyon vacuum requires that
\eq{
\tilde f > 0 , \quad \text{and} \quad h, q > 0 \,,
}
which implies that the tadpole cancellation conditions transform into,
\eq{
N_{D3}^{\text{flux}} = \tilde f h \,.
}
Thus, the tadpole is canceled by adding negatively charged objects such as orientifold planes. In this case, the non-triviality of the Galois group could imply the existence of automorphisms connecting different branches of the scalar potential. However, there are two physical vacua that are kept isolated from the other roots. The different conditions on the number of roots, the automorphisms, and the corresponding Galois Groups are shown in Figure \ref{fig:r1}.

\section{Example 2:  A model with one complex structure}
Let us consider a model in which the uplifting mechanism operates within a complex structure, incorporating non-geometric fluxes as the one studied in \cite{Blumenhagen:2015kja,Blumenhagen:2015xpa}. In this scenario, the superpotential takes the following form
\eq{
W = - i f U + i S h_0 -3 i h S U^2 - i q T,
}
with K\"ahler moduli
\eq{
K = - 3 \log \left( T + \bar T \right) - \log \left( S+ \bar S \right) -3 \log \left( U + \bar U \right).
}
In this case, the scalar potential takes the simple form where the vev's of the axionic partners are set to zero ($\rho = v = c =0$)
\eq{
V=\frac{1}{96 s \tau^3 u^3}\left( q^2 \tau^2-18 h q s \tau u^2+3 h_0^2 s^2+9 h^2 s^2 u^4-6 f h s u^3+f^2 u^2 \right) \,.
}
To obtain the univariate polynomials where the vevs are contained, we consider the derivative of the scalar potential with respect to each one of the real moduli, where we obtained a set of three polynomial equations that can be solved. \\

\subsection{AdS free tachyon}
The tachyon-free AdS physical vevs (the vevs that are real and positive) are given by
\eq{
s_0= -\frac{f}{4}\sqrt{\frac{5}{h_0 h}}, \quad \tau_0= -\frac{f}{2 q}\sqrt{\frac{5 h_0}{h}}, \quad u_0= \frac{1}{3}\sqrt{\frac{5 h_0}{h}}
}
with scalar potential at the minima given by
\eq{
V_{\text{AdS}} = -\frac{3^2}{10^2}\frac{q^3 h^2}{f^2 h}\sqrt{\frac{h}{5 h_0}}\,,
}
and the mass spectrum 
\eq{
m^2_i = \alpha_i \frac{q^3 h^2}{f^2 h_0}\sqrt{\frac{h}{5 h_0}}
}
with $i=s, u, \tau$ and where $\alpha_i$ is a positive rational number.  Now we proceed to construct  a  degree 12 polynomial $p$ on $s$ using  the Gr\"obner basis ( i.e.,  all the roots obtained by  replacing for instance $u$ and $\tau$, are contained in $p(s)$) and we observe that the SUSY minimum vevs of the saxionic partners are elements of the field extensions
\eq{
\tau_0, u_0, s_0 \in \mathbb{Q}\left( \sqrt{5 h h_0} , i  \right) 
}
for $\sqrt{ 5 h_0 h} \notin \mathbb{Q}$, which is always true since the Dirac quantization condition. This is, since $h, h_0 \in 2 \mathbb{Z}$ (we consider only even fluxes since later on we shall require the presence of orientifold planes), thus its product has to be a multiple of 2. Thus, the product of an odd number times an even number is never the square of a number, and in consequence, $\sqrt{5 h h_0}$ does not belong to $\mathbb{Q}$. \\

\begin{figure}[htbp]
   \centering
   \begin{tikzpicture}
   \draw[->,ultra thick] (-5,0)--(5,0) node[right]{Re $s$};
   \draw[->,ultra thick] (0,-3.5)--(0,3.5) node[above]{Im $s$};
      \draw[ultra thick, blue, decoration={markings, mark=at position 0.625 with {\arrow{>}}},postaction={decorate}] (-3,0) to[out=30, in = 150]  (3,0);
      \draw[ultra thick, blue, decoration={markings, mark=at position 0.625 with {\arrow{<}}},postaction={decorate}] (-3,0) to[out=-30, in = 210]  (3,0);
        \node at (-.5,1.2) {$\sigma$};
        \node at (-3.5,-0.6) {$\frac{f}{4}\sqrt{\frac{5}{h_0 h}}$};
        \node at (3.5,-0.6) {$-\frac{f}{4}\sqrt{\frac{5}{h_0 h}}$};
      \draw[dashed, red, decoration={markings, mark=at position 0.625 with {\arrow{>}}},postaction={decorate}] (0,3) to[out=210, in = 150]  (0,-3);
      \draw[dashed, red, decoration={markings, mark=at position 0.625 with {\arrow{<}}},postaction={decorate}] (0,3) to[out=-30, in = 30]  (0,-3);
        \node at (-.5,1.2) {$\sigma$};
        \node at (-1.0,3.0) {$i\frac{f}{4}\sqrt{\frac{5}{h_0 h}}$};
        \node at (-1.0,-3.5) {$-i \frac{f}{4}\sqrt{\frac{5}{h_0 h}}$};
      \draw[ultra thick,solid,red] (-2.0,-2.0) circle (0.05);
            \draw[ultra thick,solid,red] (-2.0,2.0) circle (0.05);
                  \draw[ultra thick,solid,red] (2.0,-2.0) circle (0.05);
                        \draw[ultra thick,red] (2.0,2.0) circle (0.05);

      \draw[ultra thick,solid,red] (-3.0,-1.0) circle (0.05);
            \draw[ultra thick,solid,red] (-3.0,1.0) circle (0.05);
                  \draw[ultra thick,solid,red] (3.0,-1.0) circle (0.05);
                        \draw[ultra thick,solid,red] (3.0,1.0) circle (0.05);


   \end{tikzpicture}
   
   \caption{12 Roots of p(s) for the AdS case and the trivial automorphisms. The blue lines are the automorphisms connecting the physical vacua contained in the positive part of the x-axis. The red dashed lines are the automorphism connecting the two non-physical vacua, as well as the eight red points that contain the roots of the eight-degree polynomial.}
   \label{fig:comstruct}
\end{figure}
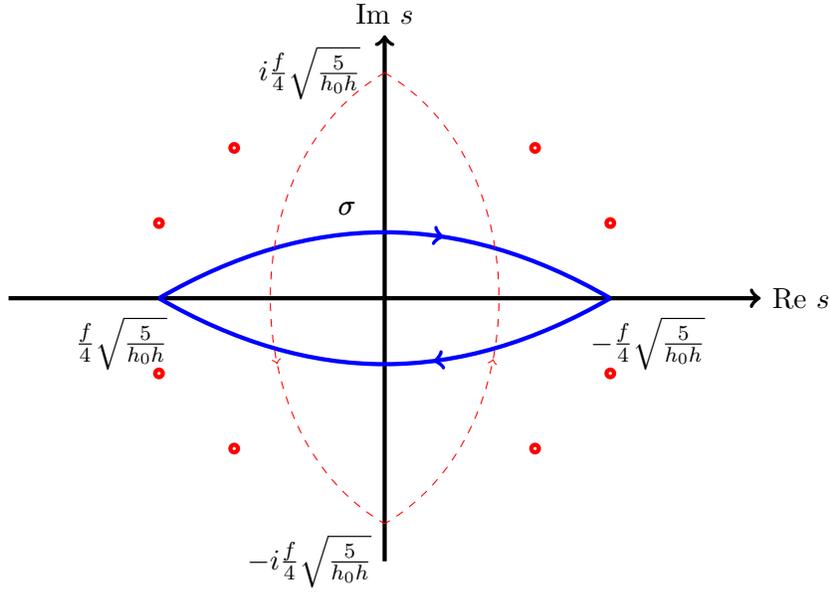

Now, let us determine the Galois group of the complete set of critical points, which are contained in $p(s)$. This degree 12 polynomial is reducible by Eisenstein criterium into two polynomials of degree 2 and one polynomial of degree 8. The degree two polynomials 
\eq{
g_1(s)= 16 h h_0 s^2 + 5 f ,\quad g_2(s)= 16 h h_0 s^2 - 5 f,
}
which can be transformed into a monic polynomial with integer coefficients
\eq{
\tilde g_1(s)= \tilde s^2 + 80 f^2 h h_0 \quad \tilde g_2(s)= \tilde s^2 - 80 f^2 h h_0
}
have $\text{disc}\, g_{1,2} = \pm 320 f^2 h h_0$, which is not a perfect square, and thus defining the Galois group of them to be $\mathbb{Z}_2$. The degree 8 polynomial
\eq{
g_3(s)=10616832 h^4 h_0^4 s^8-20295 f^4 h^2 h_0^2 s^4+125 f^8
}
can be transformed into the monic polynomial
\eq{
\tilde g_3 = \tilde s^8+ 2^{51}\cdot 3^14 \cdot 5 \cdot 11\cdot 41\cdot (f^4 h^{14} h_0^{14}) \tilde s^4+2^{119}\cdot 3^{28} \cdot 5^3 \cdot (f^8 h^{28} h^{28})
}
with discriminant $\text{disc}\, g_{3} = 2^{781} \cdot 3^{208} \cdot 5^{17} \cdot 13^4 \cdot 83^8 \cdot f^{56} \cdot h^{196} \cdot h_0^{196}$ which is not a perfect square, and thus the group $A_8$ and its subsets are excluded.  In order to determine the corresponding Galois group, one can test a chain of subgroups of $S_8$ applying the criteria given in references \cite{wildsmithcontribution}.  Following the notation for transitive groups given in \cite{transitivegroups} we find that the relevant chain of subgroups to be analyzed is given by
\eq{
\left[ S_4^2 \right] 2 \rightarrow \left[ 2^4 \right] S_4 \rightarrow \frac{1}{2} 2^4 S_4 \rightarrow {\bf \frac{1}{2} \left[ 2^4 \right] eD_4 } \rightarrow \frac{1}{2} \left[ 2^4 \right] 4 \,,
}
which has trimmed invariant polynomial over $\frac{1}{2} 2^4 S(4)$ given by\footnote{See references \cite{wildsmithcontribution}, page \cite{transitivegroups} for notation and the application of these criteria.}
\eq{
F_{D4 \rtimes D_4} = \tilde s_1 \tilde s_2 \tilde s_5 + \tilde s_1 \tilde s_6 \tilde s_5 + \tilde s_4 \tilde s_8 \tilde s_5 + \tilde s_2 \tilde s_3 \tilde s_6 + \tilde s_3 \tilde s_4 \tilde s_7 + \tilde s_2 \tilde s_6 \tilde s_7 + \tilde s_1 \tilde s_4 \tilde s_8 + \tilde s_3 \tilde s_7 \tilde s_8
}
with coset representatives of $\frac{1}{2} \left[ 2^4 \right] 4$ with respect to $\frac{1}{2} \left[ 2^4 \right] eD_4$ given by $\{ e, (2,4,6,8)\}$, yielding
\eq{
F &=-2^{43} \cdot 3^{10} \cdot 5(1+i) \sqrt[4]{2} \sqrt{3} f^3 h^{21/2} h_0^{21/2} \left(\sqrt[4]{\chi }+\sqrt[4]{\bar{\chi}}\right) \\
(2,4,6,8) F &=2^{43} \cdot 3^{10} \cdot 5(1+i) \sqrt[4]{2} \sqrt{3} f^3 h^{21/2} h_0^{21/2} \left(\sqrt[4]{\chi }+\sqrt[4]{\bar{\chi}}\right) 
}
where $\chi = 451+ 249\sqrt{39} i$. Thus, since the $F$ invariant does not contain an integer we conclude that the final Galois group of $p(s)=g_1(s)g_2(s)g_3(s)$ is a subset of 
\eq{
\text{Gal}\,(p(s)) \subset \frac{1}{2} \left[ 2^4 \right] 4\otimes \mathbb{Z}_2 \otimes \mathbb{Z}_2
} 
which is solvable and non-transitive. The roots of the defining polynomial are shown in Figure \ref{fig:comstruct}.\\

The physical vev's are contained in the field extension related to the $\mathbb{Z}_2$ subgroup of the Galois group. Thus, the fixed subgroups are composed of two elements, i.e., $\{ e, r \}$ on the physical branch. The non-physical vev's are contained in the field extension of the corresponding polynomial of degree 8 and contain a branch with two Kummer extensions of polynomials of degree four over the field extension $\mathbb{Q}(\sqrt{39})$. To define a simple extension field over $\mathbb{Q}(\sqrt{39})$, lets consider 
\eq{
\xi =\dfrac{5}{2^{18}3^{2}}\dfrac{f^4}{(h h_0)^2}\chi
}
which belongs to $\mathbb{Q}(\sqrt{39})$. Thus, only one extension field contains a vev that is real and positive for the moduli.\\

Now, let us focus on the subfields containing the physical vevs. In this case, the field extensions depend only on the NS fluxes since the vev depends linearly on the non-geometric fluxes as well as the RR fluxes. This is, a general element of the field extension $\mathbb{Q}\left( \sqrt{5 h h_0} ,  \sqrt{-5 h h_0} \right)$ can be written as,
\eq{
\alpha = a_0 + \sqrt{5 h h_0} a_1 + i a_2 + i \sqrt{5 h h_0} a_3
}
thus $\left[ \mathbb{Q}\left( \sqrt{5 h h_0} , i  \right) : \mathbb{Q} \right] = 4$ thus by the Lagrange theorem the subgroups related to the subfields have a dimension of 2 and the non-geometric as well as the RR flux dependence is contained in the coefficients $a_i$. For instance the choice $a_0 = a_2 =a_3 =0$ and $a_1 = \frac{f_1}{4 h h_0}$ gives the a vev of $s_0$. Besides, from the point of view of the Galois theory, the automorphism $\sigma_1 (\sqrt{5 h h_0}) = -\sqrt{5 h h_0}$. \\

In this case, the physical vacua require that
\eq{
f < 0, \quad \text{and} \quad h_0 ,\, h ,\, 	 q\, > 0
}
which implies that the  tadpole contribution is given by
\eq{
N_{D3} = f h
}
from which one notice that D3-branes are required. As discussed previously, the branch that contained the physical vacua connects two distinct vacua via a reflection automorphism. This implies a corresponding change in the orientation of the cycles over which  the corresponding  RR fluxes are supported on.\\

\subsection{Uplift to Minkowski}
Now, to uplift the AdS vacua to Minkowski vacua, we again consider the contribution of a $\overline{\text{D3}}$ brane given by $V_{\ov{D3}}$ with the condition of having $V=0$. The critical points are then given at
\eq{
s_0 = \frac{1}{3^{1/4}} \frac{f}{\left( 3 h h_0 \right)^{1/2}}, \quad \tau_0 =  \frac{f}{3^{1/4} q} \left( \frac{h_0}{h} \right)^{1/2}, \quad u_0 = \frac{1}{3^{1/4}}\left( \frac{h_0}{h} \right)^{1/2} 
}
which belongs to the extension $ \mathbb{Q}\left( \sqrt{3 h h_0} , 3^{1/4}  \right) $. The minimal polynomial containing all the roots is given by
\eq{
p(s) = s^4 - \frac{f^4}{27 h_0^2 h^2} = 0 \,,
}
which clearly is a non-monical polynomial. Lets consider the monic polynomial $p(\tilde s) = (27 h_0^2 h^2)^4 p\left( \frac{s}{27 h_0^2 h^2}\right)$, defined as
\eq{
\tilde s^4 - 19683 f^4 h^6 h_0^6 = 0
}
where $\tilde s = \frac{s}{27 h_0^2 h^2}$ and the roots of the modified polynomial are contained in the same splitting field as $p(s)$.\\

Thus,  the roots are contained in a degree 4 polynomial,  with integer coefficients,  which is irreducible over $\mathbb{Q}$ by the Eisenstein criterium. In consequence, the Galois group must be a transitive subgroup of $S_4$. There are five transitive subgroups of $S_4$, and since the discriminant of the defining polynomial is not a square, the Galois group shall be either $D_4$ or $\mathbb{Z}_4$. Thus, lets take test the $D_4$ invariant polynomial relative to $S_4$
\eq{
F_{D_4} = \tilde s_1 \tilde s_3 + \tilde s_2 \tilde s_4 \,,
}
for $\tilde s_i$ any root of the defining polynomial, with coset representative of $D_4$ with respect to $S_4$ as $\{ (e) , ( 23 ), ( 34 ) \}   $, yielding 
\eq{
(\text{e})F &= \tilde s_1 \tilde s_3 + \tilde s_2 \tilde s_4 = - 162 i \sqrt{3} f^2 h^3 h_0^3 \\
(23)F &= \tilde s_1 \tilde s_2 + \tilde s_3 \tilde s_4 = 162 i \sqrt{3} f^2 h^3 h_0^3 \\
(34)F &= \tilde s_1 \tilde s_4 + \tilde s_2 \tilde s_3 = 0 \\
}
since there exists one rational conjugate evaluation of $F_{D_4}$ then we conclude that the Galois group for the Minkowski vacua, is given by
\eq{
\text{Gal}\, (f(s) ) = \mathbf{D_4} \,,
} 
which is transitive and solvable\footnote{ See Appendix, section 2 for more details about the decision tree related to $S_4$.}.\\

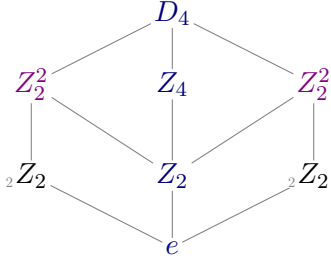
\begin{figure}[htbp]
\begin{center}
\begin{tikzpicture}[scale=1.0,sgplattice]
  \node[char] at (2,0) (1) {\gn{C1}{e}};
  \node[char] at (2,0.953) (2) {\gn{C2}{Z_2}};
  \node at (0.125,0.953) (3) {\gn{C2}{Z_2}};
  \node at (3.88,0.953) (4) {\gn{C2}{Z_2}};
  \node[norm] at (0.125,2.17) (5) {\gn{C2^2}{Z_2^2}};
  \node[char] at (2,2.17) (6) {\gn{C4}{Z_4}};
  \node[norm] at (3.88,2.17) (7) {\gn{C2^2}{Z_2^2}};
  \node[char] at (2,3.12) (8) {\gn{D4}{D_4}};
  \draw[lin] (1)--(2) (1)--(3) (1)--(4) (2)--(5) (3)--(5) (2)--(6) (2)--(7)
     (4)--(7) (5)--(8) (6)--(8) (7)--(8);
  \node[cnj=3] {2};
  \node[cnj=4] {2};
\end{tikzpicture}
\end{center}
\caption{Subgroup lattice diagram of D$_4$. The small gray index indicates that there are two copies of the group}
\label{fig:D4}
\end{figure}

In this case, the tachyon-free physical vacua require that
\eq{
f > 0,  \quad \text{and} \quad h, \, h_0, \, q \, >0 \,,
}
with a tadpole contribution given by
\eq{
N_{D3}^{\text{flux}} = f h.
}
Thus, the tadpole becomes positive, requiring O3 planes to cancel it. In this case, the subgroup lattice diagram is given by Figure \ref{fig:D4}. In this case, all the vacua are connected by automorphisms of the D$_4$ group. \\

\subsection{Uplift to dS}
\begin{figure}[htbp]
   \centering
   \begin{tikzpicture}
	 \draw[-,ultra thick] (-5,0) node[left]{AdS, $\Lambda < 0$} --(2.5,0) node[right]{12 Roots, Gal$~p(s)=G \otimes \mathbb{Z}_2 \otimes \mathbb{Z}_2$, S};
	 \draw[-,ultra thick] (-5,2) node[left]{M, $\Lambda = 0$} --(2.5,2) node[right]{4 Roots, Gal$~p(s)= D_4$, S};
	 \draw[-,ultra thick] (-5,4) node[left]{dS, $\Lambda > 0$} --(2.5,4) node[right]{34 Roots,  Gal$~p(s)= S_{34}$, NS};
	 \draw[->,ultra thick] (-4.5,0.1)  --(-4.5,1.9) node at (-4.9,1.0) {$\overline{D3}$};
	 \filldraw (1.5,4) circle (3pt) node[below]{Physical};
	 \draw[->,thick] \foreach \x in {0,1,...,3} { (-4.0+0.5*\x,0.0)  -- (-4+1.83*0,2)} ;
	 \draw[->,thick] \foreach \x in {0,1,...,6}  {(-4.0,2)  -- (-4+0.4*\x,4)} ;  \node at (-1.0,3) {\ldots};
	 \draw[->,thick,dashed] (-4.0,2)  -- (-4+0.4*9,4) ;
	 \foreach \x in {0,1,...,11} { \filldraw (-4+0.5*\x,0) circle (3pt);}  
	 \foreach \x in {0,1,2,3} { \filldraw (-4+1.83*\x,2) circle (3pt);}
	 \foreach \x in {0,1,...,6} { \filldraw (-4+0.4*\x,4) circle (3pt);}
   \end{tikzpicture}
   \caption{dS vacua are related to 34 roots of a polynomial with a Galois group given by $S_{34}$ which is a non-solvable group, contrary to the case of AdS and Minkowski. Through the lifting process, one can notice that some roots could be related and degenerated, although a much more detailed analysis is required to clearly elucidate those aspects. $G=\frac{1}{2} \left[ 2^4 \right] 4$, S= Solvable, NS= Non-Solvable.}
   \label{fig:r2}
\end{figure}
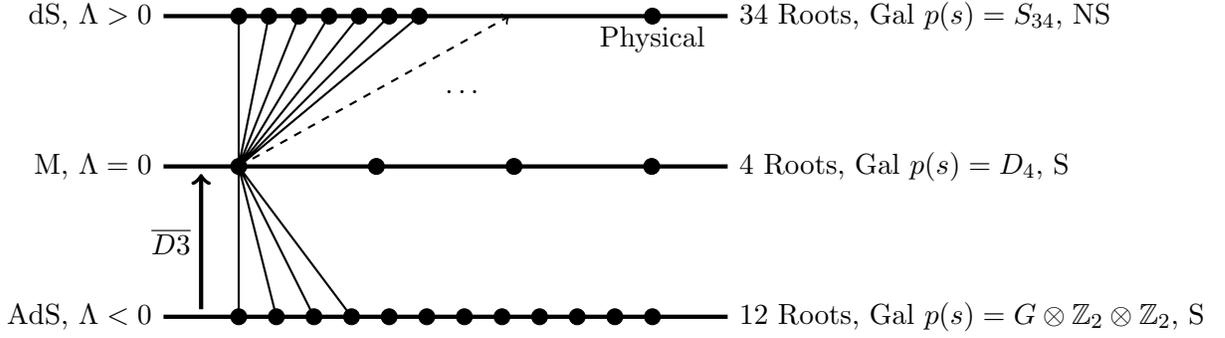

Now, let us explore the Galois group for the critical points obtained by solving the set of algebraic equations,
\eq{
\partial_u V  = \partial_\tau V = \partial_s V = 0 \,,\quad V = \Lambda
}
where $\Lambda > 0$ is the cosmological constant and the contribution to the scalar potential coming from $\ov{D3}$  given by $V_{\ov{D3}}$. In this case, the numerator $A_{\ov{D3}}$ which is related to the warping factor of the metric can be considered as a parameter that is adjusted to achieve the positive value of the vacuum and take the value
\eq{
A_{\ov{D3}} = \tau^2 \left( \Lambda - V \right)
}
thus, the derivatives of the scalar potential can be written only in terms of the moduli. In order to eliminate the variables $\tau$ and $u$ we calculate the Gr\"obner basis for the ideal generated by the derivatives of the scalar potential giving a polynomial of degree 34 for the axiodilaton, given by
\eq{
p(s) &= -f^{22} \Lambda ^2 q^{18}-512 f^{19} h_0^4 \Lambda ^3 q^{15} s^5-5109 f^{18} h^2 h_0^2 \Lambda ^2 q^{18} s^4-32 f^{17} h^4 \Lambda  q^{21} s^3 \\
&-35831808 f^{16} h_0^8 \Lambda ^4 q^{12} s^{10}+8272512 f^{15} h^2 h_0^6 \Lambda ^3 q^{15} s^9-5748880 f^{14} h^4 h_0^4 \Lambda ^2 q^{18} s^8\\
&-96672 f^{13} h^6 h_0^2 \Lambda  q^{21} s^7-18345885696 f^{13} h_0^{12} \Lambda ^5 q^9 s^{15}-2304 f^{12} h^8 q^{24} s^6\\ 
&+174439968768 f^{12} h^2 h_0^{10} \Lambda ^4 q^{12} s^{14}+17294212800 f^{11} h^4 h_0^8 \Lambda ^3 q^{15} s^{13}\\
&-251138880 f^{10} h^6 h_0^6 \Lambda ^2 q^{18} s^{12}-320979616137216 f^{10} h_0^{16} \Lambda ^6 q^6 s^{20}-6656256 f^9 h^8 h_0^4 \Lambda  q^{21} s^{11}\\
&-50008591171584 f^9 h^2 h_0^{14} \Lambda ^5 q^9 s^{19}+62208 f^8 h^{10} h_0^2 q^{24} s^{10}-37252825841664 f^8 h^4 h_0^{12} \Lambda ^4 q^{12} s^{18}\\
&-2853603758592 f^7 h^6 h_0^{10} \Lambda ^3 q^{15} s^{17}-164341563462254592 f^7 h_0^{20} \Lambda ^7 q^3 s^{25}\\
&-23145108480 f^6 h^8 h_0^8 \Lambda ^2 q^{18} s^{16}-19195373587267584 f^6 h^2 h_0^{18} \Lambda ^6 q^6 s^{24}\\
&-3418860141674496 f^5 h^4 h_0^{16} \Lambda ^5 q^9 s^{23}+2140954131336192 f^4 h^6 h_0^{14} \Lambda ^4 q^{12} s^{22}\\
&+16269814782763204608 f^3 h^2 h_0^{22} \Lambda ^7 q^3 s^{29}+2211549555185418240 f^2 h^4 h_0^{20} \Lambda ^6 q^6 s^{28}\\
&+567964443325551869952 h^2 h_0^{26} \Lambda ^8 s^{34}.
}
This polynomial is normalized to obtain a monic polynomial with integer coefficients, which is the polynomial that contains the roots of the axiodilaton at the vacuum. This polynomial is too big to be examined by trimmed invariants. However,  for specific values of the fluxes, it is possible to determine the Galois group employing the Magma package. In this case, the Galois group is the symmetric group acting on a set of cardinality 34, this is 
\eq{
\text{Gal} \left( p(s) \right) = S_{34}
}
which is non-solvable and transitive. As in the Minkowski case, the physical vacua requires the choice of orientation for the fluxes
\eq{
f > 0,  \quad \text{and} \quad h, \, h_0, \, q \, >0 \,,
}
with a tadpole contribution given by
\eq{
N_{D3}^{\text{flux}} = f h,
}
requiring O3 planes to cancel it. We show the main characteristics of the uplifting process in Figure \ref{fig:r2}.\\

%

\section{Example 3: A case with non-BPS states}
Let us consider the scalar potential generated in the presence of non-BPS states $\widehat{D5}$ and $F_5$ as considered in \cite{Damian:2022xfk}
\eq{
\tilde V = \frac{A_{\widehat{D5}}}{\sqrt{s} \tau ^{5/2}} + V
}
where
\eq{
V = \frac{A_{F_3}}{s \tau ^3}+\frac{A_{F_5}}{\tau ^4}+\frac{A_{H_3} s}{\tau ^3}+\frac{A_{O3}}{\tau ^3}.
}
The scalar potential is split into the contribution coming from the standard RR and NS fluxes, a F$_5$ flux forms coming from the torsional NS and RR fluxes warped on torsional cycles, and the contribution from orientifold 3-planes.\\

\subsection{AdS free tachyon}
In the absence of the non-BPS states A$_{\widehat{D5}}$, the scalar potential admits a stable AdS vacua obtained from the roots of the algebraic equations
\eq{
\partial_s V = \partial_\tau V = 0 \,,
}
the vev's of the saxionic fields are thus
\eq{
( s_0 , \tau ) = \left[ \left( - \sqrt{\frac{A_{F_3}}{A_{H_3}}},  \frac{4}{3}\frac{A_{F_5}}{\sqrt{4 A_{F_3} A_{H_3}}- A_{O3}} \right), 
\left( \sqrt{\frac{A_{F_3}}{A_{H_3}}},  -\frac{4}{3}\frac{A_{F_5}}{\sqrt{4 A_{F_3} A_{H_3}}+ A_{O3}} \right) \right]
}
where the physical vacua are related to the second vev.  The vacua is AdS with value
\eq{
V_{\text{min}} = -\frac{1}{3} \frac{A_{F_5}}{\tau_0^4}\,,
}
and mass spectrum 
\eq{
m^2 = \left( \frac{2 A_{H_3}^{1/2}}{s_0 \tau_0} , \frac{4 A_{F_5}}{\tau_0^6} \right)\,.
}

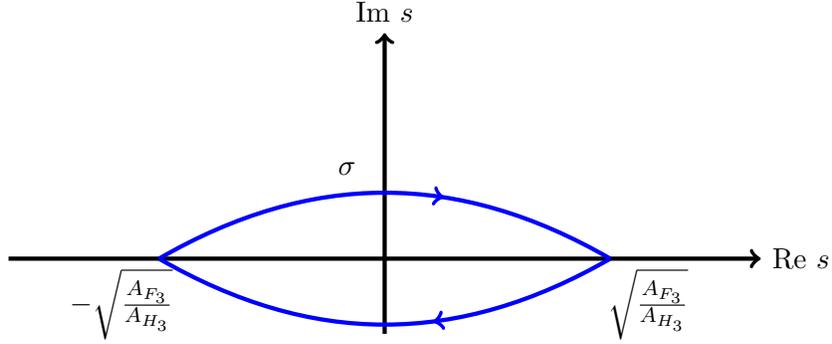
\begin{figure}[htbp]
   \centering
   \begin{tikzpicture}
   \draw[->,ultra thick] (-5,0)--(5,0) node[right]{Re $s$};
   \draw[->,ultra thick] (0,-1)--(0,3) node[above]{Im $s$};
   
   \draw[ultra thick, blue, decoration={markings, mark=at position 0.625 with {\arrow{>}}},postaction={decorate}] (-3,0) to[out=30, in = 150]  (3,0);
      \draw[ultra thick, blue, decoration={markings, mark=at position 0.625 with {\arrow{<}}},postaction={decorate}] (-3,0) to[out=-30, in = 210]  (3,0);
   
           \node at (-.5,1.2) {$\sigma$};
        \node at (-3.5,-0.6) {$-\sqrt{\frac{A_{F_3}}{A_{H_3}}}$};
  
        \node at (3.5,-0.6) {$\sqrt{\frac{A_{F_3}}{A_{H_3}}}$};
   \end{tikzpicture}
   
   \caption{Roots of p(s) and the trivial automorphisms.}
   \label{fig:nonbps}
\end{figure}

The physical vacua requires that $A_{F_3}, A_{H_3}, A_{F_5} >0$, and for the flux contribution being a rational number it is possible to conclude that the vevs are contained in the field extension
\eq{
s_0 , \tau_0 \in \mathbb{Q} \left( \sqrt {A_{F_3} A_{H_3} } \right) \,,
}
and the axiodilaton vev's are elements of the roots of a degree 2 polynomial
\begin{equation}
p(s) = s^2 - \frac{A_{F_3}}{A_{H_3}} =0 \,,
\end{equation}
this polynomial can be normalized to be a monic polynomial with integer coefficients of the form
\eq{
p(\tilde s) = \tilde s^2 -A_{F_3} A_{H_3} =0 \,,
}
with discriminant, $\Delta = 4 A_{F_3} A_{H_3}$ which could be a perfect square if $A_{F_3}  =A_{H_3}$. However, this case corresponds to a vev for the axionic field equal to 1 which could lead to an inconsistency of the tadpole cancellation conditions.  By Eisenstein criterium, the defining polynomial for $\tilde s$ is irreducible over $\mathbb{Q}$ and hence its Galois group is a transitive group, subset of the symmetric group $S_2$. The cyclic group $\mathbb{Z}_2$ is the only transitive subgroup. The splitting field is $\mathbb{Q} \left( \sqrt{\frac{A_{F_3}}{A_{H_3}}} \right)$. Thus the Galois group containing the vev of the AdS vacuum is
\eq{
\text{Gal}  ~p(s)  = \mathbb{Z}_2 \,,
}
which is transitive and solvable. The roots of the physical vacua are shown in Figure \ref{fig:nonbps}.

In this simple case the group $\mathbb{Z}_2$ contains two automorphisms related to the identity element and a reflection, this implies that the two vacua contained in the splitting field of $p(s)$ are connected by an automorphism $\sigma (s_1) = -s_2$ where $s_1$ are the two roots of the $p(s)$. However, by the action of this automorphism, the vev of the axiodilaton changes of sign making it a non-physical vacuum.\\

\subsection{Uplift to Minkowski}
Let us consider the case for $A_{\widehat{D5}} \ne 0$, and its value parametrized such that
\eq{
\tilde V= 0 \rightarrow A_{\widehat{D5}} = -\sqrt{s}\tau^{5/2} V \,,
}
the vevs for the moduli s and $\tau$ are thus contained in the roots of the algebraic equations
\eq{
\partial_s \tilde V = \partial_\tau \tilde V = 0 \,,
}
thus, eliminating $\tau$ the system is reduced to
\eq{
p(s) = 4 A_{H_3} s^2+A_{O3}s -2 A_{F_3},
}
which is normalized to the monic polynomial with integer coefficients, 
\eq{
p(\tilde s ) = \tilde s^2+A_{O3} \tilde s-8 A_{F3} A_{H3} 
}
with discriminant $\Delta = A_{O3}^2+32 A_{F3} A_{H3}$. Since the polynomial $f(\tilde s) \in \mathbb{Q}\left[ \tilde s \right]$ is of degree two, the Galois group is
\eq{
\text{Gal}\, p(s) = \mathbb{Z}_2
}
which is transitive and solvable. As in the AdS case, the non-trivial automorphism of $\mathbb{Z}_2$ implies the mapping of a physical to a non-physical vacuum.\\

\subsection{Uplift to dS: Scenario 1}
Let us consider the uplift of the AdS vacua by considering
\eq{
\tilde V= \Lambda \rightarrow A_{\widehat{D5}} = \sqrt{s}\tau^{5/2} \left( \Lambda - V \right) \,,
}
and the moduli vevs are contained at the zeros of the algebraic equations
\eq{
\partial_s V = \partial_\tau V  = 0,
}
which by eliminating $\tau$ we are left with a polynomial of degree 8 given by
\eq{
p(s) &= 16 A_{F_3}^4-80 A_{F_3}^3 A_{O3}s +16 A_{F_3}^2 \left(9 A_{O3}^2-14 A_{F_3} A_{H_3}\right) s^2 \\
&+12 A_{F_3} A_{O3} \left(68 A_{F_3} A_{H_3}-9 A_{O3}^2\right)s^3  \\
&+(1152 A_{F_3}^2 A_{H_3}^2-936 A_{F_3} A_{H_3} A_{O3}^2+256 A_{F_5}^3 \Lambda +27 A_{O3}^4)s^4 \\
&+12 A_{H_3} A_{O3} \left(27 A_{O3}^2-224 A_{F_3} A_{H_3}\right)s^5 +160 A_{H_3}^2 \left(9 A_{O3}^2-16 A_{F_3} A_{H_3}\right)s^6\\
&+2816 A_{H_3}^3 A_{O3}s^7 +2048 A_{H_3}^4 s^8,
}
which is normalized to a monic polynomial with integer coefficients with discriminant
\begin{align}
\Delta &= 2^{542} A_{F_3}^{12} A_{F_5}^{12} A_{H_3}^{180} \Lambda ^4
\bigg(1565515579392 A_{F_3}^8 A_{H_3}^8-856141332480 A_{F_3}^7 A_{H_3}^7 A_{O3}^2\\ \nonumber
&-382205952 A_{F_3}^6 A_{H_3}^6 \left(458752 A_{F_5}^3 \Lambda +23 A_{O3}^4\right)+\\ \nonumber
&5971968 A_{F_3}^5 A_{H_3}^5 A_{O3}^2 \left(10367 A_{O3}^4-69107712 A_{F_5}^3 \Lambda \right)\\\ \nonumber 
&-6912 A_{F_3}^4 A_{H_3}^4 \left(-708132732928 A_{F_5}^6 \Lambda ^2+28126089216 A_{F_5}^3 A_{O3}^4 \Lambda +401085 A_{O3}^8\right) \\
&-6912 A_{F_3}^3 A_{H_3}^3 A_{O3}^2 \left(-314136592384 A_{F_5}^6 \Lambda ^2+5659352064 A_{F_5}^3 A_{O3}^4 \Lambda +251559 A_{O3}^8\right) \\ \nonumber
&+864 A_{F_3}^2 A_{H_3}^2 \big( 962072674304 A_{F_5}^9 \Lambda ^3+436263190528 A_{F_5}^6 A_{O3}^4 \Lambda ^2 \\ \nonumber
&-4393191168 A_{F_5}^3 A_{O3}^8 \Lambda +47385 A_{O3}^{12}\big) +432 A_{F_3} A_{H_3} A_{O3}^2 \big(-1181116006400 A_{F_5}^9 \Lambda ^3 \\ \nonumber
&+56813027328 A_{F_5}^6 A_{O3}^4 \Lambda ^2-399810816 A_{F_5}^3 A_{O3}^8 \Lambda +45927 A_{O3}^{12} \big) \\ \nonumber
&+\left(256 A_{F_5}^3 \Lambda +27 A_{O3}^4\right)^2 \left(536870912 A_{F_5}^6 \Lambda ^2-3649536 A_{F_5}^3 A_{O3}^4 \Lambda +729 A_{O3}^8\right)\big).
\end{align}
First of all, observe that for $\Lambda=0$ the discriminant vanishes implying the presence of degeneracy.  Hence the two roots present in the Minkowski vacua split into different roots in the dS uplifting. Second of all, 
since $\Delta$ is not a perfect square, the Galois group cannot be a subset of $A_8$, and the decision tree for this case gives the following sequence of subgroups:
\eq{
\left[ S_4^2 \right]^2 \rightarrow \left[ 2^4 \right] S_4 \rightarrow PGL (2,7) \rightarrow {\bf {S_8}}
}
from which we conclude that the Galois group is
\eq{
\text{Gal} \, p(s) = S_8
}
which is transitive and non-solvable.\\

\subsection{Uplift to dS: Scenario 2}
\begin{figure}[htbp]
   \centering
   \begin{tikzpicture}
	 \draw[-,ultra thick] (-5,0) node[left]{AdS, $\Lambda < 0$} --(2.5,0) node[right]{2 Roots, Gal$~p(s)=\mathbb{Z}_2$, S};
	 \draw[-,ultra thick] (-5,2) node[left]{M, $\Lambda = 0$} --(2.5,2) node[right]{2 Roots, Gal$~p(s)=\mathbb{Z}_2$, S};
	 \draw[-,ultra thick] (-5,4) node[left]{dS, $\Lambda > 0$} --(2.5,4) node[right]{6/8 Roots, Gal$~p(s)=S_{6/8}$, NS};
	 \draw[->,ultra thick] (-4.5,0.1)  --(-4.5,1.9) node at (-4.9,1.0) {$\widehat{D5}$};
	  \draw[->,ultra thick] (-4.5,2.1)  --(-4.5,3.9) node at (-5.2,3.0) {$\Lambda \not = 0$};
	 \filldraw (-4+5.5/7*5,4.5) node[above]{1 physical};
	 \draw[->,thick] \foreach \x in {0,1,...,3}  {(-4.0,2)  -- (-4+5.5/7*\x,4)} ;  \node at (-1.0,3) {\ldots};
	 \draw[->,thick,dashed] (-4.0,2)  -- (-4+5.5/7*4,4) ;
	 \foreach \x in {0,1} { \filldraw (-4+5.5/1*\x,0) circle (3pt);}  
	 \foreach \x in {0,1} { \filldraw (-4+5.5*\x,2) circle (3pt);}
	 \foreach \x in {0,1,...,7} { \filldraw (-4+5.5/7*\x,4) circle (3pt);}
	 \draw[color=red,dashed] (-4+5.5/7*5,4) circle [radius=.3]; 
   \end{tikzpicture}
   \caption{In this case there are only two roots in AdS and Minkowski, and only one of them is physical. As in the previous cases, once we uplift to dS, in this case by adding non-BPS states, the degenerate roots in the Minkowski vacua split into 6 or 8 roots, fixing the Galois group to $G_{6/8}$ which is unsolvable. S = Solvable, NS = Non-solvable.}
   \label{fig:r4}
\end{figure}
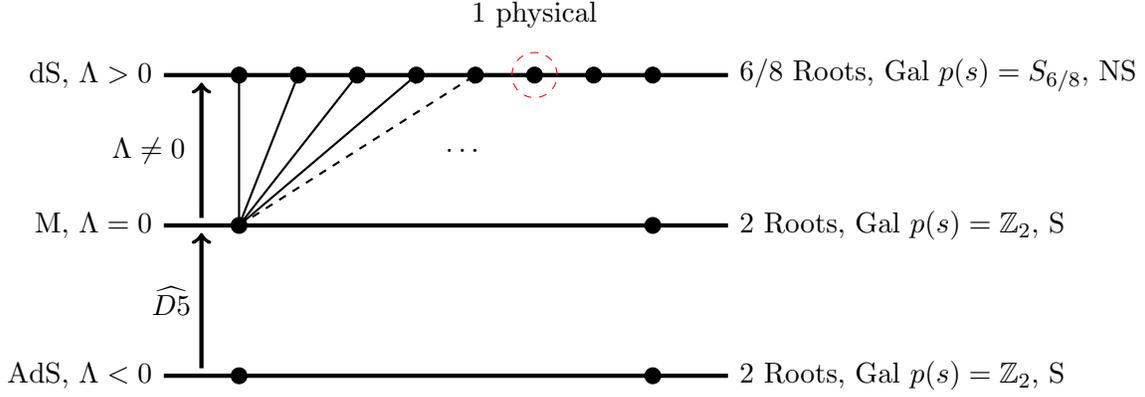

Now, let us consider the stabilization of moduli taking $A_{\widehat{D5}}$ as a free parameter and the moduli fixed by the roots of the algebraic equations
\eq{
\partial_s V = \partial_\tau V = 0 \,,
}
and eliminating $\tau$, the vev for the axiodilaton is fixed by the roots of
\eq{
p(s) &= A_{\widehat{D5}}^2 A_{F_5} s^2-2 A_{F_3}^3+12 A_{F_3}^2 A_{H_3} s^2+3 A_{F_3}^2 A_{O3} s-18 A_{F_3} A_{H_3}^2 s^4-6 A_{F_3} A_{H_3} A_{O3} s^3 \\
&+8 A_{H_3}^3 s^6+3 A_{H_3}^2 A_{O3} s^5,
}
which is normalized to the monic polynomial with integer coefficients
\eq{
p(\tilde s) &= 512 A_{\widehat{D5}}^2 A_{F_5} A_{H_3}^9 \tilde{s}^2-65536 A_{F_3}^3 A_{H_3}^{15}+12288 A_{F_3}^2 A_{H_3}^{12} A_{O3} \tilde{s}+6144 A_{F_3}^2 A_{H_3}^{10} \tilde{s}^2 \\
&-384 A_{F_3} A_{H_3}^7 A_{O3} \tilde{s}^3-144 A_{F_3} A_{H_3}^5 \tilde{s}^4+3 A_{H_3}^2 A_{O3} \tilde{s}^5+\tilde{s}^6,
}
with discriminant
\begin{align}
\Delta &= 2^{60} A_{\widehat{D5}}^4 A_{F_3}^3 A_{F_3}^2 A_{H_3}^{68} \bigg(1048576 A_{\widehat{D5}}^8 A_{F_3}^4 A_{H_3} \\ \nonumber
&+72 A_{\widehat{D5}}^6 A_{F_3}^3 \left(753664 A_{F_3}^2 A_{H_3}^2+20224 A_{F_3} A_{H_3} A_{O_3}^2-243 A_{O_3}^4\right) \\ \nonumber
& +27 A_{\widehat{D5}}^4 A_{F_3} A_{F_3}^2 \left(28098560 A_{F_3}^3 A_{H_3}^3+5960448 A_{F_3}^2 A_{H_3}^2 A_{O_3}^2-23328 A_{F_3} A_{H_3} A_{O_3}^4-729 A_{O_3}^6\right) \\ \nonumber
&+15552 A_{\widehat{D5}}^2 A_{F_3}^3 A_{F_3} A_{H_3} \left(94208 A_{F_3}^3 A_{H_3}^3+66752 A_{F_3}^2 A_{H_3}^2 A_{O_3}^2+8060 A_{F_3} A_{H_3} A_{O_3}^4-159 A_{O_3}^6\right) \\ \nonumber
&+186624 A_{F_3}^4 A_{H_3} \left(4 A_{F_3} A_{H_3}-A_{O_3}^2\right)^3 \left(64 A_{F_3} A_{H_3}+9 A_{O_3}^2\right)\bigg),
\end{align}
which is not a perfect square. Thus, the roots are contained on a polynomial of degree 6 which is irreducible over $\mathbb{Q}$. Hence the Galois group  is fixed to be
\eq{
\text{Gal} \left( p(s) \right) = S_6 \,,
}
which is transitive and non-solvable. In Figure \ref{fig:r4} we summarized the different roots of the associated polynomials as well as the corresponding Galois groups for AdS, Minkowski and dS vacua.\\

\section{Example 4: A Calabi-Yau with $h_{2,1} \ne 0$}
Let's consider the tree-level analytical models considered in \cite{Coudarchet:2022fcl}, where it is considered a non-scale type IIB compactification on a Calabi-Yau 3-fold X$_3$. As usual, the periods are written as\footnote{We are using the same notation as the reference \cite{Coudarchet:2022fcl} where the saxions are the imaginary part of the moduli, contrary to our previous examples.}
\eq{
\Pi^t = \left( \int_{B_I} \Omega, \int_{A^I} \Omega \right) \,,
}
where the cycles $\{ A^I, B_I \}$ are elements of a symplectic basis, and the complex structure is defined  as $z_i = X^i / X^0$ for $i = 1, \ldots , h^{2,1}$ together with the gauge $X^0 = 1$. Thus, the GVW superpotential is expressed as
\eq{
W= \int_{X_3} \left( F_3 - \tau H_3 \right) \wedge \Omega \,,
}
which is written as
\eq{
W = -\frac{1}{6} N_A^0 \kappa_{ijk} z^i z^j z^k +\frac{1}{2} \kappa_{ijk} N_A^i z^j z^k + \left( N_A^j a_{ij} + N_i^B-N_A^0 c_i \right) z^i - \kappa_0 N^0_A- N_A^i c_i + N_0^B \,.
}
where the flux vector is defined as
\eq{
N=\left(
\begin{matrix}
f_0^B \\
f_i^B \\
f_A^0\\
f_A^i\\
\end{matrix}
\right)
-
\tau
\left(
\begin{matrix}
h_0^B \\
h_i^B \\
h_A^0\\
h_A^i\\
\end{matrix}
\right)
=
\left(
\begin{matrix}
N_0^B \\
N_i^B \\
N_A^0\\
N_A^i\\
\end{matrix}
\right)
}
which allows writing the tadpole as
\eq{
N_{\text{flux}} = \frac{N^\dagger \cdot \Sigma \cdot \Pi}{\tau - \bar \tau} \,.
}
On the other hand, the K\"ahler potential is given by 
\eq{
K= -2 \ln \mathcal{V} - \ln \left( -i (\tau - \bar \tau) \right) - \ln \left(  -i \Pi^\dagger \cdot \Sigma \cdot \Pi \right) 
}
where $\mathcal{V}$ is the volume modulus, $\tau = b^0+i t^0$ is the axio-dilaton and the complex structures are $z^i=b^i + i t^i$. The canonical symplectic matrix is
\eq{
\Sigma = \left(
\begin{matrix}
0 & \mathbf{1} \\
-\mathbf{1} & 0 \\
\end{matrix}
\right)
}
which explicitly give us the K\"ahler potential
\eq{
K= -2 \ln \mathcal{V} - \ln \left( 2 t^0 \right) -\ln \left( \frac{4}{3} \kappa_{ijk} t^i t^j t^k-2 \text{Im} \kappa_0 \right) \,,
}
In \cite{Coudarchet:2022fcl}, the authors focus on the dubbed IIB1 scenario, which is based upon selecting the appropriate RR and NS fluxes that makes the superpotential able to be written in the bilinear form
\eq{ \label{eq:supbl}
W = \frac{1}{2} \vec Z^t M \vec Z + \vec L \cdot \vec Z + Q
}
which comes from an F-theory compactification in the large complex structure region. This scenario allows writing the scalar potential into a bilinear structure which is quite useful to express the critical point of the scalar potential in order to systematically study the corresponding vacua. In this scenario, the following fluxes are set to zero,
\eq{
f_A^0 = h_A^0 = h_A^i = 0 \quad i \in \{1, \ldots , h^{2,1} \}
}
reducing the tadpole to
\eq{
N_{\text{flux}} = - f_A^i h^B_i \,,
}
where
\eq{
M = \left(
\begin{matrix}
0 & - \vec (h^{B})^T  \\
-\vec h^B & h_{ij} \\
\end{matrix} \,,
\right) \quad 
\vec L = \left( -h_0^B , f_i^B + a_{ij} f^j_A \right) \,, \quad 
Q = f_0^B - c_i f^i_A \,.
}
The scalar potential constructed from Eq. \ref{eq:supbl} admits non-supersymmetric vacua contained in the set of algebraic equations
\eq{ \label{eq:vac}
\tilde x^2 &= \frac{N_{\text{flux}}}{\kappa^H} \hat x \left( \hat x -1 \right) \,, \\
x^0 &= \frac{\mathcal{S}\hat x (\hat x-1 )}{N_{\text{flux}}} + \tilde x - 2 \hat x \tilde x \,, \\
\left( 2 \hat x^3 -3 \hat x^2 + \alpha \right)^2 &= 16 \frac{N^3_{\text{flux}}}{\mathcal{S} \kappa^H} \hat x^3 \left( \hat x -1 \right)^3 \,. 
 }
where
\eq{
x^0 &= \frac{4}{3} \frac{Q'}{\text{Im} \kappa_0} t^0 \,, \quad x^i = \frac{4}{3} \frac{Q'}{\text{Im}\, \kappa_0} t^i \,, \quad \alpha = \frac{2^5 Q'^3}{3^2 \left(\text{Im}\, \kappa_0 \right)^2 \mathcal{S}}\\
\mathcal{S} &= \kappa_{ijk} f_A^i f_A^j f_A^k \,, \quad \kappa^H = \kappa_{ijk} S^{il} h_l^B S^{jm}h_m^B S^{kn}h_n^B 
}
and 
\eq{
x^i = \hat x f^i_A + \tilde x S^{ij} h_i^B \,,
}
thus by solving Eq. \ref{eq:vac} it is possible to obtain the vev's of the saxionic moduli. To look for the Galois group we transform the non-monic polynomial into the monic polynomial with integer coefficients.  As it is observed, the roots of $\hat x$ are contained on a degree 6 polynomial
\eq{
p(\hat x) =\kappa^H \mathcal{S} \left(\alpha +2 \hat x^3-3 \hat x^2\right)^2-16N_{\text{flux}}^3 (\hat x-1)^3 \hat x^3 \,,
}
thus, there are 16 transitive groups of degree 6, with 5 of them being non-solvable. However, in this case, the discriminant is
\begin{align}
\Delta &= 2^{62}\cdot 3^6 \cdot N_{\text{flux}}^9 \mathcal{S}^4 (\alpha -1)^4 \alpha ^4 (\kappa^H)^4 \left(4 N_{\text{flux}}^3-\mathcal{S} \kappa^H \right)^5 \left(\mathcal{S} \kappa^H-4 N_{\text{flux}}^3\right)^{16} \\ \nonumber
& \left(N_{\text{flux}}^6+4 \mathcal{S} (\alpha -1) \alpha  \kappa^H N_{\text{flux}}^3-\mathcal{S}^2 (\alpha -1) \alpha  (\kappa^H)^2\right)
\end{align}
discarding the alternating group $A_6$ as well as its subgroups as the Galois group. The decision tree goes as follows
\eq{
\left[ S_{3}^2 \right] 2 \rightarrow 2^3 \left[ S_3 \right] \rightarrow \sqrt{\nabla ( f(\hat x))} \in \mathbb{Z} \rightarrow PGL (2,5)  \rightarrow \mathbf{S}_6
}
yielding the Galois group
\eq{
\text{Gal} \left( p(\hat x) \right) = S_6 \,,
}
which is non-solvable and transitive. 

\section{Conclusions}
We have chosen four different scenarios in which it is possible, at least numerically, to obtain stable de Sitter (dS) vacua by adding exotic structures such as non-geometric fluxes or non-BPS branes, resulting in a polynomial scalar potential. This potential is crucial to our study as we are interested in computing the corresponding Galois group, which provides important information about the algebraic structure of the roots of the polynomial representing the scalar potential's moduli. To compute the Galois group of a specific modulus, we utilized conventional techniques reported in existing literature, although the analysis can be extended to other moduli yielding similar outcomes.\\

In all four selected examples, the presence of dS critical points is consistently associated with the absence of analyticity caused by the unsolvability of the Galois group, in contrast to Anti-de Sitter (AdS) and Minkowski vacua. The solvability of the Galois group implies the possibility of expressing the polynomial roots in terms of radicals and fractional numbers. Notably, for all uplifted dS vacua, the Galois group becomes unsolvable. This observation suggests a connection between the challenge of finding analytical examples of dS vacua and the unsolvability of their associated Galois groups.\\

Additionally, although the Galois group is non-solvable, we expect an automorphism connecting different dS critical points since it relates to the critical points associated with all vacua. However, in all cases, the physical vacua are either isolated from the other critical points or connected to a non-physical vacuum characterized by a negative value of the saxionic components' vacuum expectation value (vev). We also observe that a change in the orientations of fluxes can be interpreted as the result of $\mathbb{Z}_2$ automorphisms, which act as a bridge between the automorphisms of Minkowski vacua. As a Minkowski vacuum continuously deforms into a dS vacuum, its roots undergo degeneration, leading to a similar degeneration in a branch of physical vacua and resulting in changes in vacuum energy.\\

Therefore, based on our studied examples, it is expected that dS vacua in perturbative compactifications, if they exist, are associated with a non-solvable Galois group. If the swampland de Sitter conjecture holds true, one could conclude that all vacua in the landscape are linked to a solvable Galois group, at least within a perturbative framework. A more detailed study is undoubtedly required to confirm this association.\\

\section{Appendix}
\subsection{A (slightly more formal) brief introduction to Galois theory}

For those who want to read a little more formal description of the steps to compute the Galois group, in this section, we provide more specific details. A reader not interested in these details can safely skip this appendix. For a more detailed description, the interested reader must check \cite{brzezinski2018galois,stewart2022galois}. 
Overall, computing the Galois group of a given polynomial can be a complex and involved process that requires knowledge of algebraic structures, group theory, and field theory.\\

The Galois theory gives a relation between the roots of a polynomial $p(x)$ and the group action of the splitting fields related to such roots.  The splitting field of a polynomial $p(x)$, denoted  $L[p]$ is defined as the smallest field $\mathbb{Q}$ generated by the roots $\alpha_i$ of $p(s)$ and is denoted as,
\begin{equation}
L[p] = \mathbb{Q} \left( \alpha_1, \alpha_2, \ldots \right),
\end{equation}

The relation between $L \left[ p \right]$ and $\{ \alpha_i \}$  is codified in the so-called group of automorphisms $\text{Aut}[L]$ of $L[p]$, which is constructed by bijective function $\sigma : L \rightarrow L$ such that, for all $\alpha, \beta \in L$,
\begin{itemize}
\item $\sigma ( \alpha + \beta ) = \sigma (\alpha) + \sigma (\beta)$,
\item $\sigma ( \alpha \beta) = \sigma (\alpha) \sigma (\beta)$,
\item $\sigma (\alpha) = \alpha \quad \forall \alpha \in K\subset L$,
\end{itemize}
where the third property defines the so-called K-automorphism. Thus, the group of all K-automorphisms of L is denoted as $\text{Gal}(L/K)$ and is known as the Galois group of L over K. Notice that if $\alpha_i$ is a root of $p$, thus
\eq{
p(\sigma (\alpha_i ) ) = \sigma (p ( \alpha_i )) = 0 \,.
}
and similarly $\sigma^{-1} \left( \alpha_i \right)$ is also a root of $p$, allowing us to abuse the notation and refer to Gal$(L/K )$ as the Galois group of the polynomial $p(x)$ denoted by Gal $~p(x)$.\\

In order to compute the Galois group of the splitting field $L \left[ p \right]$, it is required for the latter to be normal and separable. If an irreducible polynomial contains a root in $L$, it must contain all the roots.  Separability of an extension field implies that each element on it is separable and each element is algebraic and is the root of a separable polynomial $f \in \mathbb{Q} (x)$.  The separability of a polynomial can be verified if $f (\alpha)$ and its formal derivative $f' (\alpha)$ have no trivial factors in common.\\

An important invariant to classify the Galois group is the discriminant, which is defined as follows. Let $f \in \mathbb{Q} (x)$, be a degree n polynomial with roots $\alpha_i$, the discriminant is defined as
\eq{
\text{disc}\, f = \prod_{i<j}^n \left( \alpha_j-\alpha_i \right)^2 \,.
}
Thus, $\text{disc}\, f(x) $ is nonzero if $f(x)$ is separable (no repeated roots). Thus, for a separable polynomial $f$, $\text{disc}\, f $ is a symmetric polynomial in the the roots $\alpha_i$, and in consequence is fixed by the Galois group of $f(x)$.\\

The discriminant of $f(x)$ is useful in reducing the possible Galois groups, for instance, if $\text{disc}\, f $ is a square in $\mathbb{Q}$ then the Galois group of $f(x)$ is a subset of the alternating group $A_n$. For instance, in the case of an irreducible degree 3 polynomial the Galois group is uniquely determined by this criterium since there are only two transitive subgroups of $S_3$.

Now, let us define the fixed subfields under the K-automorphisms as
\eq{
L^H = \{ x \in L : \forall_{\sigma \in H} \sigma (x) = x \}
}
as well as the fixed elements of the Galois group
\eq{
G(L/M) = \{ \sigma \in G(L/K) : \forall_{x\in M} \sigma (x) =x \}
} 
where the Galois group of an irreducible polynomial over the rationals of degree n is a transitive subgroup of the symmetric group $\text{S}_n$. Namely, for any two roots of the polynomial f, there exists an element of the Galois group such that $\gamma( a) = b$. Thus, this restriction on the possible subgroups reduces considerably the possibilities to take into account. We shall employ the Eisenstein criterium to test the irreducibility of the polynomials over $\mathbb{Q}$. For this criterium, let $f(x) = \sum_{i=1}^n a_i x^i$ be a polynomial over $\mathbb{Z}$, the polynomial shall be irreducible over $\mathbb{Q}$ if there is a prime q such that i) $q$ does not divide $a_n$, ii) $q^2$ does not divide $a_0$ and iii) $q$ divides all $a_i$ for $i<n$.\\

To look for the Galois group, in the following, we shall employ the algorithm proposed by Stauduhar \cite{stauduhar1973determination}, which is adequate for monic irreducible polynomials over  $\mathbb{Z}$. This is not a limitation of the algorithm even when the polynomials that are typically involved in a tree-level vacuum in string theory are non-monic polynomials over $\mathbb{Z}$, due to the flux quantization conditions. This is, since any non-monic polynomial can be easily transformed into a monic polynomial by taking the transformation
\begin{equation}
f (\phi ) \rightarrow a_n^n f \left( \frac{\phi}{a}\right) \,,
\end{equation}
where $a_n$ is the coefficient associated with the element of a higher power in the polynomial and $f( \phi) \in \mathbb{Q} \left[ \phi \right]$ preserves the splitting field. The Galois group is computed through invariants derived from the structure of the groups. In the cases studied below, we consider the G-relative H-invariants given that $H < G < \text{S}_n$, thus a polynomial $F$ is G-relative H-invariant if its stabilizer
\begin{equation}
\text{Stab}_G\, F = \{ \sigma \in G | F^{\sigma} = F \}
\end{equation}
equals H. Furthermore, if $\text{Gal}(f) \leq G $ where $\text{Gal}(f)$ acts on the roots $\alpha_i$, the relative resolvent is given by
\begin{equation}
R_F(y) = \prod_{\sigma\in G//H} \left( y -  F^\sigma(\alpha_1, \alpha_2, \ldots , \alpha_n )\right)
\end{equation}
where $G//H$ is the set of representatives of right cosets $H\sigma${\footnote{To compute the right cosets of $H \subset G$, it is selected an element of $g \in G$ which is not contained in $H$ and $H g$ is computed for each element.}} and 
\begin{equation}
F = \sum_{\sigma \in G} x_{\sigma(1)}x^2_{\sigma(2)} \ldots x_{\sigma(n)}^n \,.
\end{equation}
is a $S_n$-relative H-invariant polynomial. However, these polynomials are usually computationally too expensive, since each term needs $n-2$ multiplications, thus in the following the trimmed invariants proposed by \cite{GEISSLER2000653} shall be employed.\\

The computation of the Galois group is based upon identifying the transitive subgroup of $\text{S}_n$ by traversing the lattice of subgroups down to the Galois group of the polynomial. To check the lattice subgroups, we shall employ de decision tree together with the trimmed invariant polynomials proposed in \cite{wildsmithcontribution}. In this manner, the Galois group is found independently of the numerical value for the fluxes. Besides, the Galois group is checked against a specific selection of fluxes and the classification provided by Magma. \\

Magma calculates the Galois group through the algorithm proposed by Kl\"uners and Geissler, which goes as follows.   For an irreducible polynomial $f$ the algorithm starts by factoring  $f$ modulo several primes $p$ to look for a prime such that the least common multiple of the factors is not quite large keeping $f$ square-free. If the Galois group is in $\text{S}_n$ or $\text{A}_n$, the group can be detected from the degrees of the modulo p-factors. If the Galois group is not $\text{S}_n$ or $\text{A}_n$ the next step is to calculate the roots of the polynomial up to high precision. Numerically, this step is carried on by employing taking an approximation of the roots in the p-adic field, however, for the present paper, the roots of the polynomial can be written in an explicit manner depending on the NS or RR fluxes for the case of solvable Galois groups.\\

Once the roots are known, the next steps require the knowledge of a suitable starting group to start the Stauduhar iteration. An obvious starting group is $\text{S}_n$, however, this starting group is not always the best choice, since the computation of the $\text{S}
_n$-relative H-invariants could become computationally quite expensive. Instead, to determine the starting group, it is better to compute all subfields of the stem field $E = \mathbb{Q}\left[x \right]/f$ of $f$, the corresponding block systems and the largest transitive group G admitting those blocks.\\

The third step is the iteration of the Stauduhar test. As a result, we have a chain of subgroups with the properties described above; hence, we apply those techniques to verify the result.\\

On the other hand, for reducible polynomials the Galois group of a compositum polynomial $f = f_i f_j$ and for Galois groups $G_1 (f_1)$ and $G_2 (f_2)$, the total Galois group is isomorphic to a group $G \in G_1 \times G_2$ \cite{awtrey2017determining,SUTHERLAND201573}. Besides, the Galois groups of the factors shall be invariant of $G$.\\

%
%

\subsection{Decision trees}
The decision tree for $S_4$ \cite {wildsmithcontribution}  (for more complicated decision trees check the reference). On the decision tree, the arrows at the left indicate the case where the answer is positive. In consequence, the dashed lines imply only subgroups of $A_4$.\\
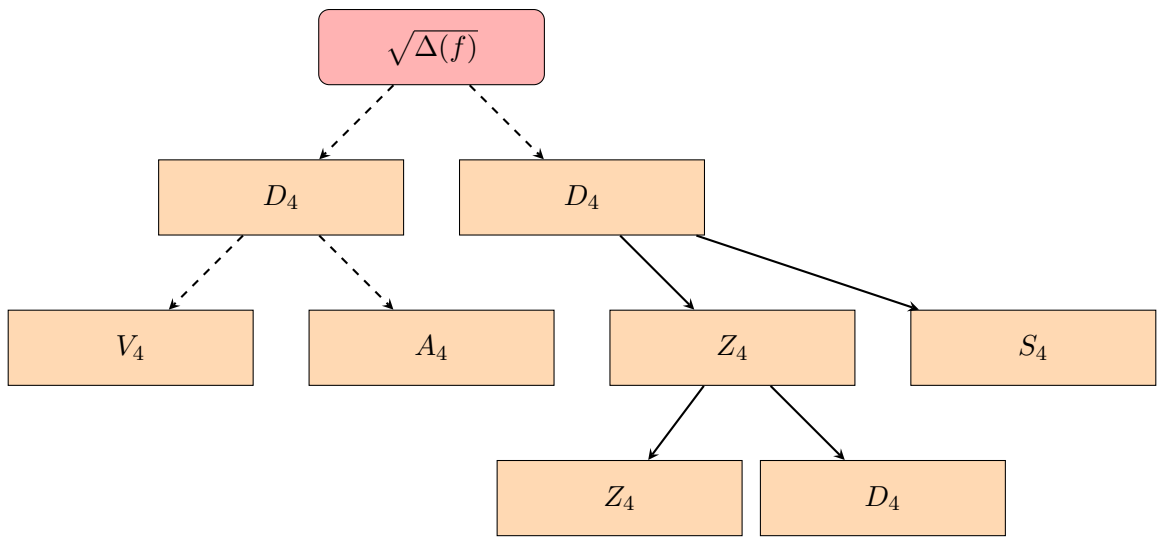
\begin{figure}[!htbp]
\begin{center}
\begin{tikzpicture}[node distance=2cm]

\node (S4) [startstop] {$\sqrt{\Delta (f)}$};
\node (D4) [process, below of=S4,xshift=-2cm] {$D_4$};
\node (D4r) [process, right of=D4,xshift=2cm] {$D_4$};

\node (V4) [process, below of=D4,xshift=-2cm] {$V_4$};
\node (A4) [process, below of=D4,xshift=+2cm] {$A_4$};
\node (Z4) [process, right of=A4,xshift=2cm] {$Z_4$};
\node (S4b) [process, right of=Z4,xshift=2cm] {$S_4$};

\node (Z4b) [process, below of=Z4,xshift=-1.5cm] {$Z_4$};
\node (D4b) [process, right of=Z4b,xshift=+1.5cm] {$D_4$};

\draw [arrow,dashed] (S4) -- (D4);
\draw [arrow,dashed] (S4) -- (D4r);
\draw [arrow,dashed] (D4) -- (V4);
\draw [arrow,dashed] (D4) -- (A4);
\draw [arrow] (D4r) -- (Z4);
\draw [arrow] (D4r) -- (S4b);
\draw [arrow] (Z4) -- (Z4b);
\draw [arrow] (Z4) -- (D4b);

\end{tikzpicture}
\end{center}
\caption{\label{fig:DS4} Decision tree for  $S_4$}
\end{figure}

For this particular case, the search for the Galois group starts by checking if the discriminant is a perfect square. If it is not a perfect square, then the decision tree leads to check if the invariant polynomial corresponding to the subgroup $D_4$ contains any integer root, in case the invariant polynomial contains an integer root, then the roots are orderer and the invariant polynomial for the $\mathbb{Z}_4$ subgroup is checked for invariant roots, and the algorithm finishes by checking the invariant polynomial of either $\mathbb{Z}_4$ or $D_4$ depending if the invariant polynomial contains or not an integer value. 

\bibliographystyle{JHEP}

\providecommand{\href}[2]{#2}\begingroup\raggedright\begin{thebibliography}{10}

\bibitem{Palti:2019pca}
E.~Palti, \emph{{The Swampland: Introduction and Review}},
  \href{https://doi.org/10.1002/prop.201900037}{\emph{Fortsch. Phys.}
  {\bfseries 67} (2019) 1900037}
  [\href{https://arxiv.org/abs/1903.06239}{{\ttfamily 1903.06239}}].

\bibitem{vanBeest:2021lhn}
M.~van Beest, J.~Calder\'on-Infante, D.~Mirfendereski and I.~Valenzuela,
  \emph{{Lectures on the Swampland Program in String Compactifications}},
  \href{https://doi.org/10.1016/j.physrep.2022.09.002}{\emph{Phys. Rept.}
  {\bfseries 989} (2022) 1} [\href{https://arxiv.org/abs/2102.01111}{{\ttfamily
  2102.01111}}].

\bibitem{Grana:2021zvf}
M.~Gra\~na and A.~Herr\'aez, \emph{{The Swampland Conjectures: A Bridge from
  Quantum Gravity to Particle Physics}},
  \href{https://doi.org/10.3390/universe7080273}{\emph{Universe} {\bfseries 7}
  (2021) 273} [\href{https://arxiv.org/abs/2107.00087}{{\ttfamily
  2107.00087}}].

\bibitem{Agmon:2022thq}
N.~B. Agmon, A.~Bedroya, M.~J. Kang and C.~Vafa, \emph{{Lectures on the string
  landscape and the Swampland}},
  \href{https://arxiv.org/abs/2212.06187}{{\ttfamily 2212.06187}}.

\bibitem{Garg:2018reu}
S.~K. Garg and C.~Krishnan, \emph{{Bounds on Slow Roll and the de Sitter
  Swampland}}, \href{https://doi.org/10.1007/JHEP11(2019)075}{\emph{JHEP}
  {\bfseries 11} (2019) 075}
  [\href{https://arxiv.org/abs/1807.05193}{{\ttfamily 1807.05193}}].

\bibitem{Ooguri:2018wrx}
H.~Ooguri, E.~Palti, G.~Shiu and C.~Vafa, \emph{{Distance and de Sitter
  Conjectures on the Swampland}},
  \href{https://doi.org/10.1016/j.physletb.2018.11.018}{\emph{Phys. Lett. B}
  {\bfseries 788} (2019) 180}
  [\href{https://arxiv.org/abs/1810.05506}{{\ttfamily 1810.05506}}].

\bibitem{Andriot:2018mav}
D.~Andriot and C.~Roupec, \emph{{Further refining the de Sitter swampland
  conjecture}}, \href{https://doi.org/10.1002/prop.201800105}{\emph{Fortsch.
  Phys.} {\bfseries 67} (2019) 1800105}
  [\href{https://arxiv.org/abs/1811.08889}{{\ttfamily 1811.08889}}].

\bibitem{Blumenhagen:2019kqm}
R.~Blumenhagen, M.~Brinkmann and A.~Makridou, \emph{{A Note on the dS Swampland
  Conjecture, Non-BPS Branes and K-Theory}},
  \href{https://doi.org/10.1002/prop.201900068}{\emph{Fortsch. Phys.}
  {\bfseries 67} (2019) 1900068}
  [\href{https://arxiv.org/abs/1906.06078}{{\ttfamily 1906.06078}}].

\bibitem{Seo:2019mfk}
M.-S. Seo, \emph{{Thermodynamic interpretation of the de Sitter swampland
  conjecture}},
  \href{https://doi.org/10.1016/j.physletb.2019.134904}{\emph{Phys. Lett. B}
  {\bfseries 797} (2019) 134904}
  [\href{https://arxiv.org/abs/1907.12142}{{\ttfamily 1907.12142}}].

\bibitem{Brandenberger:2020oav}
R.~Brandenberger, V.~Kamali and R.~O. Ramos, \emph{{Strengthening the de Sitter
  swampland conjecture in warm inflation}},
  \href{https://doi.org/10.1007/JHEP08(2020)127}{\emph{JHEP} {\bfseries 08}
  (2020) 127} [\href{https://arxiv.org/abs/2002.04925}{{\ttfamily
  2002.04925}}].

\bibitem{Damian2023some}
C.~Damian and O.~Loaiza-Brito, \emph{Some remarks on swampland conjectures,
  fluxes and k-theory in iib toroidal compactifications}, {\emph{Annals of
  Physics} (2023) 169334}.

\bibitem{Leontaris:2023obe}
G.~K. Leontaris and P.~Shukla, \emph{{Seeking de-Sitter Vacua in the String
  Landscape}},  3, 2023, \href{https://arxiv.org/abs/2303.16689}{{\ttfamily
  2303.16689}}.

\bibitem{Damian:2013dq}
C.~Damian, L.~R. Diaz-Barron, O.~Loaiza-Brito and M.~Sabido, \emph{{Slow-Roll
  Inflation in Non-geometric Flux Compactification}},
  \href{https://doi.org/10.1007/JHEP06(2013)109}{\emph{JHEP} {\bfseries 06}
  (2013) 109} [\href{https://arxiv.org/abs/1302.0529}{{\ttfamily 1302.0529}}].

\bibitem{Damian:2013dwa}
C.~Damian and O.~Loaiza-Brito, \emph{{More stable de Sitter vacua from S-dual
  nongeometric fluxes}},
  \href{https://doi.org/10.1103/PhysRevD.88.046008}{\emph{Phys. Rev. D}
  {\bfseries 88} (2013) 046008}
  [\href{https://arxiv.org/abs/1304.0792}{{\ttfamily 1304.0792}}].

\bibitem{Blumenhagen:2015kja}
R.~Blumenhagen, A.~Font, M.~Fuchs, D.~Herschmann, E.~Plauschinn, Y.~Sekiguchi
  et~al., \emph{{A Flux-Scaling Scenario for High-Scale Moduli Stabilization in
  String Theory}},
  \href{https://doi.org/10.1016/j.nuclphysb.2015.06.003}{\emph{Nucl. Phys. B}
  {\bfseries 897} (2015) 500}
  [\href{https://arxiv.org/abs/1503.07634}{{\ttfamily 1503.07634}}].

\bibitem{Damian:2018tlf}
C.~Damian and O.~Loaiza-Brito, \emph{{Two-Field Axion Inflation and the
  Swampland Constraint in the Flux-Scaling Scenario}},
  \href{https://doi.org/10.1002/prop.201800072}{\emph{Fortsch. Phys.}
  {\bfseries 67} (2019) 1800072}
  [\href{https://arxiv.org/abs/1808.03397}{{\ttfamily 1808.03397}}].

\bibitem{Plauschinn:2018wbo}
E.~Plauschinn, \emph{{Non-geometric backgrounds in string theory}},
  \href{https://doi.org/10.1016/j.physrep.2018.12.002}{\emph{Phys. Rept.}
  {\bfseries 798} (2019) 1} [\href{https://arxiv.org/abs/1811.11203}{{\ttfamily
  1811.11203}}].

\bibitem{Andriot:2019wrs}
D.~Andriot, \emph{{Open problems on classical de Sitter solutions}},
  \href{https://doi.org/10.1002/prop.201900026}{\emph{Fortsch. Phys.}
  {\bfseries 67} (2019) 1900026}
  [\href{https://arxiv.org/abs/1902.10093}{{\ttfamily 1902.10093}}].

\bibitem{Shukla:2019wfo}
P.~Shukla, \emph{{Dictionary for the type II nongeometric flux
  compactifications}},
  \href{https://doi.org/10.1103/PhysRevD.103.086009}{\emph{Phys. Rev. D}
  {\bfseries 103} (2021) 086009}
  [\href{https://arxiv.org/abs/1909.07391}{{\ttfamily 1909.07391}}].

\bibitem{Shukla:2022srx}
P.~Shukla, \emph{{On stable type IIA de-Sitter vacua with geometric flux}},
  \href{https://doi.org/10.1140/epjc/s10052-023-11361-w}{\emph{Eur. Phys. J. C}
  {\bfseries 83} (2023) 196}
  [\href{https://arxiv.org/abs/2202.12840}{{\ttfamily 2202.12840}}].

\bibitem{Coudarchet:2022fcl}
T.~Coudarchet, F.~Marchesano, D.~Prieto and M.~A. Urkiola, \emph{{Analytics of
  type IIB flux vacua and their mass spectra}},
  \href{https://doi.org/10.1007/JHEP01(2023)152}{\emph{JHEP} {\bfseries 01}
  (2023) 152} [\href{https://arxiv.org/abs/2212.02533}{{\ttfamily
  2212.02533}}].

\bibitem{awtrey2017determining}
C.~Awtrey, T.~Cesarski and P.~Jakes, \emph{Determining galois groups of
  reducible polynomials via discriminants and linear resolvents}, {\emph{JP
  Journal of Algebra Number Theory and Applications} {\bfseries 39} (2017)
  685}.

\bibitem{wildsmithcontribution}
T.~Wildsmith, \emph{A contribution to the computation of galois groups},
  \href{https://doi.org/https://doi.org/10.13140/RG .2 .2 .18201
  .24161}{\emph{Thesis for: Master of Research} (2019) }.

\bibitem{stauduhar1973determination}
R.~P. Stauduhar, \emph{The determination of galois groups}, {\emph{Mathematics
  of computation} {\bfseries 27} (1973) 981}.

\bibitem{GEISSLER2000653}
K.~Geissler and J.~Klüners, \emph{Galois group computation for rational
  polynomials},
  \href{https://doi.org/https://doi.org/10.1006/jsco.2000.0377}{\emph{Journal
  of Symbolic Computation} {\bfseries 30} (2000) 653}.

\bibitem{SUTHERLAND201573}
N.~Sutherland, \emph{Computing galois groups of polynomials (especially over
  function fields of prime characteristic)},
  \href{https://doi.org/https://doi.org/10.1016/j.jsc.2014.09.043}{\emph{Journal
  of Symbolic Computation} {\bfseries 71} (2015) 73}.

\bibitem{stewart2022galois}
I.~Stewart, \emph{Galois theory}. CRC press, 2022.

\bibitem{transitivegroups}
T.~U.~K. Universit\"at~Paderborn, \emph{A database for number fields},  2013.

\bibitem{Blumenhagen:2015xpa}
R.~Blumenhagen, C.~Damian, A.~Font, D.~Herschmann and R.~Sun, \emph{{The
  Flux-Scaling Scenario: De Sitter Uplift and Axion Inflation}},
  \href{https://doi.org/10.1002/prop.201600030}{\emph{Fortsch. Phys.}
  {\bfseries 64} (2016) 536}
  [\href{https://arxiv.org/abs/1510.01522}{{\ttfamily 1510.01522}}].

\bibitem{Damian:2022xfk}
C.~Damian and O.~Loaiza-Brito, \emph{{Metastable vacua from torsion and machine
  learning}}, \href{https://doi.org/10.1140/epjc/s10052-022-11118-x}{\emph{Eur.
  Phys. J. C} {\bfseries 82} (2022) 1129}
  [\href{https://arxiv.org/abs/2205.12373}{{\ttfamily 2205.12373}}].

\bibitem{brzezinski2018galois}
J.~Brzezi{\'n}ski, \emph{Galois theory through exercises}. Springer, 2018.

\end{thebibliography}\endgroup

\providecommand{\href}[2]{#2}\begingroup\raggedright\endgroup

\end{document}